\documentclass[a4paper,12pt]{article}
\usepackage{amsfonts}
\usepackage{amsmath}
\usepackage{amssymb}

\makeatletter
 
\def\diagram{\m@th\leftwidth=\z@ \rightwidth=\z@ \topheight=\z@
\botheight=\z@ \setbox\@picbox\hbox\bgroup}
 
\def\enddiagram{\egroup\wd\@picbox\rightwidth\unitlength
\ht\@picbox\topheight\unitlength \dp\@picbox\botheight\unitlength
\hskip\leftwidth\unitlength\box\@picbox}
 
\def\bfig{\begin{diagram}}
\def\efig{\end{diagram}}
\newcount\wideness \newcount\leftwidth \newcount\rightwidth
\newcount\highness \newcount\topheight \newcount\botheight
 
\def\ratchet#1#2{\ifnum#1<#2 \global #1=#2 \fi}
 
\def\putbox(#1,#2)#3{%
\horsize{\wideness}{#3} \divide\wideness by 2
{\advance\wideness by #1 \ratchet{\rightwidth}{\wideness}}
{\advance\wideness by -#1 \ratchet{\leftwidth}{\wideness}}
\vertsize{\highness}{#3} \divide\highness by 2
{\advance\highness by #2 \ratchet{\topheight}{\highness}}
{\advance\highness by -#2 \ratchet{\botheight}{\highness}}
\put(#1,#2){\makebox(0,0){$#3$}}}
 
\def\putlbox(#1,#2)#3{%
\horsize{\wideness}{#3}
{\advance\wideness by #1 \ratchet{\rightwidth}{\wideness}}
{\ratchet{\leftwidth}{-#1}}
\vertsize{\highness}{#3} \divide\highness by 2
{\advance\highness by #2 \ratchet{\topheight}{\highness}}
{\advance\highness by -#2 \ratchet{\botheight}{\highness}}
\put(#1,#2){\makebox(0,0)[l]{$#3$}}}
 
\def\putrbox(#1,#2)#3{%
\horsize{\wideness}{#3}
{\ratchet{\rightwidth}{#1}}
{\advance\wideness by -#1 \ratchet{\leftwidth}{\wideness}}
\vertsize{\highness}{#3} \divide\highness by 2
{\advance\highness by #2 \ratchet{\topheight}{\highness}}
{\advance\highness by -#2 \ratchet{\botheight}{\highness}}
\put(#1,#2){\makebox(0,0)[r]{$#3$}}}

\def\adjust[#1]{} % For compatibility
 
\newcount \coefa
\newcount \coefb
\newcount \coefc
\newcount\tempcounta
\newcount\tempcountb
\newcount\tempcountc
\newcount\tempcountd
\newcount\xext
\newcount\yext
\newcount\xoff
\newcount\yoff
\newcount\gap%
\newcount\arrowtypea
\newcount\arrowtypeb
\newcount\arrowtypec
\newcount\arrowtyped
\newcount\arrowtypee
\newcount\height
\newcount\width
\newcount\xpos
\newcount\ypos
\newcount\run
\newcount\rise
\newcount\arrowlength
\newcount\halflength
\newcount\arrowtype
\newdimen\tempdimen
\newdimen\xlen
\newdimen\ylen
\newsavebox{\tempboxa}%
\newsavebox{\tempboxb}%
\newsavebox{\tempboxc}%
 
\newdimen\w@dth
 
\def\setw@dth#1#2{\setbox\z@\hbox{\m@th$#1$}\w@dth=\wd\z@
\setbox\@ne\hbox{\m@th$#2$}\ifnum\w@dth<\wd\@ne \w@dth=\wd\@ne \fi
\advance\w@dth by 1.2em}
 
\def\t@^#1_#2{\allowbreak\def\n@one{#1}\def\n@two{#2}\mathrel
{\setw@dth{#1}{#2}
\mathop{\hbox to \w@dth{\rightarrowfill}}\limits
\ifx\n@one\empty\else ^{\box\z@}\fi
\ifx\n@two\empty\else _{\box\@ne}\fi}}
\def\t@@^#1{\@ifnextchar_{\t@^{#1}}{\t@^{#1}_{}}}
\def\to{\@ifnextchar^{\t@@}{\t@@^{}}}
 
\def\t@left^#1_#2{\def\n@one{#1}\def\n@two{#2}\mathrel{\setw@dth{#1}{#2}
\mathop{\hbox to \w@dth{\leftarrowfill}}\limits
\ifx\n@one\empty\else ^{\box\z@}\fi
\ifx\n@two\empty\else _{\box\@ne}\fi}}
\def\t@@left^#1{\@ifnextchar_{\t@left^{#1}}{\t@left^{#1}_{}}}
\def\toleft{\@ifnextchar^{\t@@left}{\t@@left^{}}}
 
\def\two@^#1_#2{\allowbreak
\def\n@one{#1}\def\n@two{#2}\mathrel{\setw@dth{#1}{#2}
\mathop{\vcenter{\lineskip\z@\baselineskip\z@
                 \hbox to \w@dth{\rightarrowfill}%
                 \hbox to \w@dth{\rightarrowfill}}%
       }\limits
\ifx\n@one\empty\else ^{\box\z@}\fi
\ifx\n@two\empty\else _{\box\@ne}\fi}}
\def\tw@@^#1{\@ifnextchar _{\two@^{#1}}{\two@^{#1}_{}}}
\def\two{\@ifnextchar ^{\tw@@}{\tw@@^{}}}
 
\def\tofr@^#1_#2{\def\n@one{#1}\def\n@two{#2}\mathrel{\setw@dth{#1}{#2}
\mathop{\vcenter{\hbox to \w@dth{\rightarrowfill}\kern-1.7ex
                 \hbox to \w@dth{\leftarrowfill}}%
       }\limits
\ifx\n@one\empty\else ^{\box\z@}\fi
\ifx\n@two\empty\else _{\box\@ne}\fi}}
\def\t@fr@^#1{\@ifnextchar_ {\tofr@^{#1}}{\tofr@^{#1}_{}}}
\def\tofro{\@ifnextchar^ {\t@fr@}{\t@fr@^{}}}

\def\mon{\mathop{\m@th\hbox to
      14.6\P@{\lasyb\char'51\hskip-2.1\P@$\arrext$\hss
$\mathord\rightarrow$}}\limits} % width of \epi
\def\leftmono{\mathrel{\m@th\hbox to
14.6\P@{$\mathord\leftarrow$\hss$\arrext$\hskip-2.1\P@\lasyb\char'50%
}}\limits} % width of \epi
\mathchardef\arrext="0200       % amr minus for arrow extension (see \into)

\def\settypes(#1,#2,#3){\arrowtypea#1 \arrowtypeb#2 \arrowtypec#3}
\def\settoheight#1#2{\setbox\@tempboxa\hbox{#2}#1\ht\@tempboxa\relax}%
\def\settodepth#1#2{\setbox\@tempboxa\hbox{#2}#1\dp\@tempboxa\relax}%
\def\settokens`#1`#2`#3`#4`{%
     \def\tokena{#1}\def\tokenb{#2}\def\tokenc{#3}\def\tokend{#4}}
\def\setsqparms[#1`#2`#3`#4;#5`#6]{%
\arrowtypea #1
\arrowtypeb #2
\arrowtypec #3
\arrowtyped #4
\width #5
\height #6
}
\def\setpos(#1,#2){\xpos=#1 \ypos#2}

\def\settriparms[#1`#2`#3;#4]{\settripairparms[#1`#2`#3`1`1;#4]}%
 
\def\settripairparms[#1`#2`#3`#4`#5;#6]{%
\arrowtypea #1
\arrowtypeb #2
\arrowtypec #3
\arrowtyped #4
\arrowtypee #5
\width #6
\height #6
}
 
\def\resetparms{\settripairparms[1`1`1`1`1;500]\width 500}%default values%
 
\resetparms
 
\def\mvector(#1,#2)#3{%%
\put(0,0){\vector(#1,#2){#3}}%
\put(0,0){\vector(#1,#2){26}}%
}
\def\evector(#1,#2)#3{{%%
\arrowlength #3
\put(0,0){\vector(#1,#2){\arrowlength}}%
\advance \arrowlength by-30
\put(0,0){\vector(#1,#2){\arrowlength}}%
}}
 
\def\horsize#1#2{%
\settowidth{\tempdimen}{$#2$}%
#1=\tempdimen
\divide #1 by\unitlength
}
 
\def\vertsize#1#2{%
\settoheight{\tempdimen}{$#2$}%
#1=\tempdimen
\settodepth{\tempdimen}{$#2$}%
\advance #1 by\tempdimen
\divide #1 by\unitlength
}
 
\def\putvector(#1,#2)(#3,#4)#5#6{{%
\ifnum3<\arrowtype
\putdashvector(#1,#2)(#3,#4)#5\arrowtype
\else
\ifnum\arrowtype<-3
\putdashvector(#1,#2)(#3,#4)#5\arrowtype
\else
\xpos=#1
\ypos=#2
\run=#3
\rise=#4
\arrowlength=#5
\ifnum \arrowtype<0
    \ifnum \run=0
        \advance \ypos by-\arrowlength
    \else
        \tempcounta \arrowlength
        \multiply \tempcounta by\rise
        \divide \tempcounta by\run
        \ifnum\run>0
            \advance \xpos by\arrowlength
            \advance \ypos by\tempcounta
        \else
            \advance \xpos by-\arrowlength
            \advance \ypos by-\tempcounta
        \fi
    \fi
    \multiply \arrowtype by-1
    \multiply \rise by-1
    \multiply \run by-1
\fi
\ifcase \arrowtype
\or \put(\xpos,\ypos){\vector(\run,\rise){\arrowlength}}%
\or \put(\xpos,\ypos){\mvector(\run,\rise)\arrowlength}%
\or \put(\xpos,\ypos){\evector(\run,\rise){\arrowlength}}%
\fi\fi\fi
}}
 
\def\putsplitvector(#1,#2)#3#4{%%
\xpos #1
\ypos #2
\arrowtype #4
\halflength #3
\arrowlength #3
\gap 140
\advance \halflength by-\gap
\divide \halflength by2
\ifnum\arrowtype>0
   \ifcase \arrowtype
   \or \put(\xpos,\ypos){\line(0,-1){\halflength}}%
       \advance\ypos by-\halflength
       \advance\ypos by-\gap
       \put(\xpos,\ypos){\vector(0,-1){\halflength}}%
   \or \put(\xpos,\ypos){\line(0,-1)\halflength}%
       \put(\xpos,\ypos){\vector(0,-1)3}%
       \advance\ypos by-\halflength
       \advance\ypos by-\gap
       \put(\xpos,\ypos){\vector(0,-1){\halflength}}%
   \or \put(\xpos,\ypos){\line(0,-1)\halflength}%
       \advance\ypos by-\halflength
       \advance\ypos by-\gap
       \put(\xpos,\ypos){\evector(0,-1){\halflength}}%
   \fi
\else \arrowtype=-\arrowtype
   \ifcase\arrowtype
   \or \advance \ypos by-\arrowlength
       \put(\xpos,\ypos){\line(0,1){\halflength}}%
       \advance\ypos by\halflength
       \advance\ypos by\gap
       \put(\xpos,\ypos){\vector(0,1){\halflength}}%
   \or \advance \ypos by-\arrowlength
       \put(\xpos,\ypos){\line(0,1)\halflength}%
       \put(\xpos,\ypos){\vector(0,1)3}%
       \advance\ypos by\halflength
       \advance\ypos by\gap
       \put(\xpos,\ypos){\vector(0,1){\halflength}}%
   \or \advance \ypos by-\arrowlength
       \put(\xpos,\ypos){\line(0,1)\halflength}%
       \advance\ypos by\halflength
       \advance\ypos by\gap
       \put(\xpos,\ypos){\evector(0,1){\halflength}}%
   \fi
\fi
}
 
\def\putmorphism(#1)(#2,#3)[#4`#5`#6]#7#8#9{{%
\run #2
\rise #3
\ifnum\rise=0
  \puthmorphism(#1)[#4`#5`#6]{#7}{#8}#9%
\else\ifnum\run=0
  \putvmorphism(#1)[#4`#5`#6]{#7}{#8}#9%
\else
\setpos(#1)%
\arrowlength #7
\arrowtype #8
\ifnum\run=0
\else\ifnum\rise=0
\else
\ifnum\run>0
    \coefa=1
\else
   \coefa=-1
\fi
\ifnum\arrowtype>0
   \coefb=0
   \coefc=-1
\else
   \coefb=\coefa
   \coefc=1
   \arrowtype=-\arrowtype
\fi
\width=2
\multiply \width by\run
\divide \width by\rise
\ifnum \width<0  \width=-\width\fi
\advance\width by60
\if l#9 \width=-\width\fi
\putbox(\xpos,\ypos){#4}%            %node 1
{\multiply \coefa by\arrowlength%      %node 2
\advance\xpos by\coefa
\multiply \coefa by\rise
\divide \coefa by\run
\advance \ypos by\coefa
\putbox(\xpos,\ypos){#5} }%
{\multiply \coefa by\arrowlength%      %label
\divide \coefa by2
\advance \xpos by\coefa
\advance \xpos by\width
\multiply \coefa by\rise
\divide \coefa by\run
\advance \ypos by\coefa
\if l#9%
   \putrbox(\xpos,\ypos){#6}%
\else\if r#9%
   \putlbox(\xpos,\ypos){#6}%
\fi\fi }%
{\multiply \rise by-\coefc%             %arrow
\multiply \run by-\coefc
\multiply \coefb by\arrowlength
\advance \xpos by\coefb
\multiply \coefb by\rise
\divide \coefb by\run
\advance \ypos by\coefb
\multiply \coefc by70
\advance \ypos by\coefc
\multiply \coefc by\run
\divide \coefc by\rise
\advance \xpos by\coefc
\multiply \coefa by140
\multiply \coefa by\run
\divide \coefa by\rise
\advance \arrowlength by\coefa
\ifcase\arrowtype
\or \put(\xpos,\ypos){\vector(\run,\rise){\arrowlength}}%
\or \put(\xpos,\ypos){\mvector(\run,\rise){\arrowlength}}%
\or \put(\xpos,\ypos){\evector(\run,\rise){\arrowlength}}%
\fi}\fi\fi\fi\fi}}

\newcount\numbdashes \newcount\lengthdash \newcount\increment
 
\def\howmanydashes{% Actually returns both number and length
\numbdashes=\arrowlength \lengthdash=40
\divide\numbdashes by \lengthdash
\lengthdash=\arrowlength
\divide\lengthdash by \numbdashes
\increment=\lengthdash
\multiply\lengthdash by 3
\divide\lengthdash by 5
}
 
\def\putdashvector(#1)(#2,#3)#4#5{%
\ifnum#3=0 \putdashhvector(#1){#4}#5
\else
\ifnum#2=0
\putdashvvector(#1){#4}#5\fi\fi}
 
\def\putdashhvector(#1,#2)#3#4{{%
\arrowlength=#3 \howmanydashes
\multiput(#1,#2)(\increment,0){\numbdashes}%
{\vrule height .4pt width \lengthdash\unitlength}
\arrowtype=#4 \xpos=#1
\ifnum\arrowtype<0 \advance\arrowtype by 7 \fi
\ifcase\arrowtype
\or \advance\xpos by 10
    \put(\xpos,#2){\vector(-1,0){\lengthdash}}
    \advance\xpos by 40
    \put(\xpos,#2){\vector(-1,0){\lengthdash}}
\or \advance \xpos by 10
    \put(\xpos,#2){\vector(-1,0){\lengthdash}}
    \advance\xpos by  \arrowlength
    \advance\xpos by  -50
    \put(\xpos,#2){\vector(-1,0){\lengthdash}}
\or \advance\xpos by 10
    \put(\xpos,#2){\vector(-1,0){\lengthdash}}
\or \advance\xpos by \arrowlength
    \advance\xpos by -\lengthdash
    \put(\xpos,#2){\vector(1,0){\lengthdash}}
\or {\advance\xpos by 10
    \put(\xpos,#2){\vector(1,0){\lengthdash}}}
    \advance\xpos by \arrowlength
    \advance\xpos by -\lengthdash
    \put(\xpos,#2){\vector(1,0){\lengthdash}}
\or \advance\xpos by \arrowlength
    \advance\xpos by -\lengthdash
    \put(\xpos,#2){\vector(1,0){\lengthdash}}
    \advance\xpos by -40
    \put(\xpos,#2){\vector(1,0){\lengthdash}}
   \fi
}}
 
\def\putdashvvector(#1,#2)#3#4{{%
\arrowlength=#3 \howmanydashes
\ypos=#2 \advance\ypos by -\arrowlength
\multiput(#1,#2)(0,\increment){\numbdashes}%
    {\vrule width .4pt height \lengthdash\unitlength}
\arrowtype=#4 \ypos=#2
\ifnum\arrowtype<0 \advance\arrowtype by 7 \fi
\ifcase\arrowtype
\or \advance\ypos by \arrowlength \advance\ypos by -40
    \put(#1,\ypos){\vector(0,1){\lengthdash}}
    \advance\ypos by -40
    \put(#1,\ypos){\vector(0,1){\lengthdash}}
\or \advance\ypos by 10
    \put(#1,\ypos){\vector(0,1){\lengthdash}}
    \advance\ypos by \arrowlength \advance\ypos by -40
    \put(#1,\ypos){\vector(0,1){\lengthdash}}
\or \advance\ypos by \arrowlength \advance\ypos by -40
    \put(#1,\ypos){\vector(0,1){\lengthdash}}
\or \advance\ypos by 10
    \put(#1,\ypos){\vector(0,-1){\lengthdash}}
\or \advance\ypos by 10
    \put(#1,\ypos){\vector(0,-1){\lengthdash}}
    \advance\ypos by \arrowlength \advance\ypos by -40
    \put(#1,\ypos){\vector(0,-1){\lengthdash}}
\or \advance\ypos by 10
    \put(#1,\ypos){\vector(0,-1){\lengthdash}}
    \advance\ypos by 40
    \put(#1,\ypos){\vector(0,-1){\lengthdash}}
\fi
}}
 
\def\puthmorphism(#1,#2)[#3`#4`#5]#6#7#8{{%
\xpos #1
\ypos #2
\width #6
\arrowlength #6
\arrowtype=#7
\putbox(\xpos,\ypos){#3\vphantom{#4}}%
{\advance \xpos by\arrowlength
\putbox(\xpos,\ypos){\vphantom{#3}#4}}%
\horsize{\tempcounta}{#3}%
\horsize{\tempcountb}{#4}%
\divide \tempcounta by2
\divide \tempcountb by2
\advance \tempcounta by30
\advance \tempcountb by30
\advance \xpos by\tempcounta
\advance \arrowlength by-\tempcounta
\advance \arrowlength by-\tempcountb
\putvector(\xpos,\ypos)(1,0)\arrowlength\arrowtype
\divide \arrowlength by2
\advance \xpos by\arrowlength
\vertsize{\tempcounta}{#5}%
\divide\tempcounta by2
\advance \tempcounta by20
\if a#8 %
   \advance \ypos by\tempcounta
   \putbox(\xpos,\ypos){#5}%
\else
   \advance \ypos by-\tempcounta
   \putbox(\xpos,\ypos){#5}%
\fi}}
 
\def\putvmorphism(#1,#2)[#3`#4`#5]#6#7#8{{%
\xpos #1
\ypos #2
\arrowlength #6
\arrowtype #7
\settowidth{\xlen}{$#5$}%
\putbox(\xpos,\ypos){#3}%
{\advance \ypos by-\arrowlength
\putbox(\xpos,\ypos){#4}}%
{\advance\arrowlength by-140
\advance \ypos by-70
\ifdim\xlen>0pt
   \if m#8%
      \putsplitvector(\xpos,\ypos)\arrowlength\arrowtype
   \else
   \putvector(\xpos,\ypos)(0,-1)\arrowlength\arrowtype
   \fi
\else
   \putvector(\xpos,\ypos)(0,-1)\arrowlength\arrowtype
\fi}%
\ifdim\xlen>0pt
   \divide \arrowlength by2
   \advance\ypos by-\arrowlength
   \if l#8%
      \advance \xpos by-40
      \putrbox(\xpos,\ypos){#5}%
   \else\if r#8%
      \advance \xpos by40
      \putlbox(\xpos,\ypos){#5}%
   \else
      \putbox(\xpos,\ypos){#5}%
   \fi\fi
\fi
}}
 
\def\putsquarep<#1>(#2)[#3;#4`#5`#6`#7]{{%
\setsqparms[#1]%
\setpos(#2)%
\settokens`#3`%
\puthmorphism(\xpos,\ypos)[\tokenc`\tokend`{#7}]{\width}{\arrowtyped}b%
\advance\ypos by \height
\puthmorphism(\xpos,\ypos)[\tokena`\tokenb`{#4}]{\width}{\arrowtypea}a%
\putvmorphism(\xpos,\ypos)[``{#5}]{\height}{\arrowtypeb}l%
\advance\xpos by \width
\putvmorphism(\xpos,\ypos)[``{#6}]{\height}{\arrowtypec}r%
}}
 
\def\putsquare{\@ifnextchar <{\putsquarep}{\putsquarep%
   <\arrowtypea`\arrowtypeb`\arrowtypec`\arrowtyped;\width`\height>}}
\def\square{\@ifnextchar< {\squarep}{\squarep
   <\arrowtypea`\arrowtypeb`\arrowtypec`\arrowtyped;\width`\height>}}
\def\squarep<#1>[#2`#3`#4`#5;#6`#7`#8`#9]{{%       %     #2------>#3
\setsqparms[#1]%                                   %      |       |
\diagram%                                          %      |       |
\putsquarep<\arrowtypea`\arrowtypeb`\arrowtypec`%  %    #7|       |#8
\arrowtyped;\width`\height>%                       %      |       |
(0,0)[#2`#3`#4`{#5};#6`#7`#8`{#9}]%                %      |       |
\enddiagram%                                       %      v       v
}}                                                 %     #4------>#5
\def\putptrianglep<#1>(#2,#3)[#4`#5`#6;#7`#8`#9]{{%
\settriparms[#1]%
\xpos=#2 \ypos=#3
\advance\ypos by \height
\puthmorphism(\xpos,\ypos)[#4`#5`{#7}]{\height}{\arrowtypea}a%
\putvmorphism(\xpos,\ypos)[`#6`{#8}]{\height}{\arrowtypeb}l%
\advance\xpos by\height
\putmorphism(\xpos,\ypos)(-1,-1)[``{#9}]{\height}{\arrowtypec}r%
}}
 
\def\putptriangle{\@ifnextchar <{\putptrianglep}{\putptrianglep
   <\arrowtypea`\arrowtypeb`\arrowtypec;\height>}}
\def\ptriangle{\@ifnextchar <{\ptrianglep}{\ptrianglep
   <\arrowtypea`\arrowtypeb`\arrowtypec;\height>}}
\def\ptrianglep<#1>[#2`#3`#4;#5`#6`#7]{{%%    %      #2----->#3
\settriparms[#1]%                             %      |      /
\diagram%                                     %      |     /
\putptrianglep<\arrowtypea`\arrowtypeb`%      %    #6|    /#7
\arrowtypec;\height>%                         %      |   /
(0,0)[#2`#3`#4;#5`#6`{#7}]%                   %      |  /
\enddiagram%%                                 %      v v
}}                                            %      #4
 
\def\putqtrianglep<#1>(#2,#3)[#4`#5`#6;#7`#8`#9]{{%
\settriparms[#1]%
\xpos=#2 \ypos=#3
\advance\ypos by\height
\puthmorphism(\xpos,\ypos)[#4`#5`{#7}]{\height}{\arrowtypea}a%
\putmorphism(\xpos,\ypos)(1,-1)[``{#8}]{\height}{\arrowtypeb}l%
\advance\xpos by\height
\putvmorphism(\xpos,\ypos)[`#6`{#9}]{\height}{\arrowtypec}r%
}}
 
\def\putqtriangle{\@ifnextchar <{\putqtrianglep}{\putqtrianglep
   <\arrowtypea`\arrowtypeb`\arrowtypec;\height>}}
\def\qtriangle{\@ifnextchar <{\qtrianglep}{\qtrianglep
   <\arrowtypea`\arrowtypeb`\arrowtypec;\height>}}
\def\qtrianglep<#1>[#2`#3`#4;#5`#6`#7]{{%%    %        #2----->#3
\settriparms[#1]%                             %         \      |
\width=\height                                %          \     |
\diagram%                                     %         #6\    |#7
\putqtrianglep<\arrowtypea`\arrowtypeb`%      %            \   |
\arrowtypec;\height>%                         %             \  |
(0,0)[#2`#3`#4;#5`#6`{#7}]%                   %              v v
\enddiagram%%                                 %               #4
}}
 
\def\putdtrianglep<#1>(#2,#3)[#4`#5`#6;#7`#8`#9]{{%
\settriparms[#1]%
\xpos=#2 \ypos=#3
\puthmorphism(\xpos,\ypos)[#5`#6`{#9}]{\height}{\arrowtypec}b%
\advance\xpos by \height \advance\ypos by\height
\putmorphism(\xpos,\ypos)(-1,-1)[``{#7}]{\height}{\arrowtypea}l%
\putvmorphism(\xpos,\ypos)[#4``{#8}]{\height}{\arrowtypeb}r%
}}
 
\def\putdtriangle{\@ifnextchar <{\putdtrianglep}{\putdtrianglep
   <\arrowtypea`\arrowtypeb`\arrowtypec;\height>}}
\def\dtriangle{\@ifnextchar <{\dtrianglep}{\dtrianglep
   <\arrowtypea`\arrowtypeb`\arrowtypec;\height>}}
\def\dtrianglep<#1>[#2`#3`#4;#5`#6`#7]{{%%    %                  / |
\settriparms[#1]%                             %                 /  |
\width=\height                                %              #5/   |#6
\diagram%                                     %               /    |
\putdtrianglep<\arrowtypea`\arrowtypeb`%      %              /     |
\arrowtypec;\height>%                         %             v      v
(0,0)[#2`#3`#4;#5`#6`{#7}]%                   %            #3----->#4
\enddiagram%%                                 %                #7
}}
 
\def\putbtrianglep<#1>(#2,#3)[#4`#5`#6;#7`#8`#9]{{%
\settriparms[#1]%
\xpos=#2 \ypos=#3
\puthmorphism(\xpos,\ypos)[#5`#6`{#9}]{\height}{\arrowtypec}b%
\advance\ypos by\height
\putmorphism(\xpos,\ypos)(1,-1)[``{#8}]{\height}{\arrowtypeb}r%
\putvmorphism(\xpos,\ypos)[#4``{#7}]{\height}{\arrowtypea}l%
}}
 
\def\putbtriangle{\@ifnextchar <{\putbtrianglep}{\putbtrianglep
   <\arrowtypea`\arrowtypeb`\arrowtypec;\height>}}
\def\btriangle{\@ifnextchar <{\btrianglep}{\btrianglep
   <\arrowtypea`\arrowtypeb`\arrowtypec;\height>}}
\def\btrianglep<#1>[#2`#3`#4;#5`#6`#7]{{%%   %              | \
\settriparms[#1]%                            %              |  \
\width=\height                               %            #5|   \#6
\diagram%                                    %              |    \
\putbtrianglep<\arrowtypea`\arrowtypeb`%     %              |     \
\arrowtypec;\height>%                        %              v      v
(0,0)[#2`#3`#4;#5`#6`{#7}]%                  %              #3----->#4
\enddiagram%%                                %                 #7
}}
 
\def\putAtrianglep<#1>(#2,#3)[#4`#5`#6;#7`#8`#9]{{%
\settriparms[#1]%
\xpos=#2 \ypos=#3
{\multiply \height by2
\puthmorphism(\xpos,\ypos)[#5`#6`{#9}]{\height}{\arrowtypec}b}%
\advance\xpos by\height \advance\ypos by\height
\putmorphism(\xpos,\ypos)(-1,-1)[#4``{#7}]{\height}{\arrowtypea}l%
\putmorphism(\xpos,\ypos)(1,-1)[``{#8}]{\height}{\arrowtypeb}r%
}}
 
\def\putAtriangle{\@ifnextchar <{\putAtrianglep}{\putAtrianglep
   <\arrowtypea`\arrowtypeb`\arrowtypec;\height>}}
\def\Atriangle{\@ifnextchar <{\Atrianglep}{\Atrianglep
   <\arrowtypea`\arrowtypeb`\arrowtypec;\height>}}
\def\Atrianglep<#1>[#2`#3`#4;#5`#6`#7]{{%%         %         /   \
\settriparms[#1]%                                  %        /     \
\width=\height                                     %     #5/       \#6
\diagram%                                          %      /         \
\putAtrianglep<\arrowtypea`\arrowtypeb`%           %     /           \
\arrowtypec;\height>%                              %    v             v
(0,0)[#2`#3`#4;#5`#6`{#7}]%                        %   #3------------>#4
\enddiagram%%                                      %          #7
}}
 
\def\putAtrianglepairp<#1>(#2)[#3;#4`#5`#6`#7`#8]{{%
\settripairparms[#1]%
\setpos(#2)%
\settokens`#3`%
\puthmorphism(\xpos,\ypos)[\tokenb`\tokenc`{#7}]{\height}{\arrowtyped}b%
\advance\xpos by\height
\puthmorphism(\xpos,\ypos)[\phantom{\tokenc}`\tokend`{#8}]%
{\height}{\arrowtypee}b%
\advance\ypos by\height
\putmorphism(\xpos,\ypos)(-1,-1)[\tokena``{#4}]{\height}{\arrowtypea}l%
\putvmorphism(\xpos,\ypos)[``{#5}]{\height}{\arrowtypeb}m%
\putmorphism(\xpos,\ypos)(1,-1)[``{#6}]{\height}{\arrowtypec}r%
}}
 
\def\putAtrianglepair{\@ifnextchar <{\putAtrianglepairp}{\putAtrianglepairp%
   <\arrowtypea`\arrowtypeb`\arrowtypec`\arrowtyped`\arrowtypee;\height>}}
\def\Atrianglepair{\@ifnextchar <{\Atrianglepairp}{\Atrianglepairp%
   <\arrowtypea`\arrowtypeb`\arrowtypec`\arrowtyped`\arrowtypee;\height>}}
 
\def\Atrianglepairp<#1>[#2;#3`#4`#5`#6`#7]{{%           %  #2a
\settripairparms[#1]%                         %           / | \
\settokens`#2`%                               %          /  |  \
\width=\height                                %       #3/  #4   \#5
\diagram%                                     %        /    |    \
\putAtrianglepairp                            %       /     |     \
<\arrowtypea`\arrowtypeb`\arrowtypec`%        %      v      v      v
\arrowtyped`\arrowtypee;\height>%             %     #2b---->#2c---->#2d
(0,0)[{#2};#3`#4`#5`#6`{#7}]%                 %         #6     #7
\enddiagram%%
}}
 
\def\putVtrianglep<#1>(#2,#3)[#4`#5`#6;#7`#8`#9]{{%
\settriparms[#1]%
\xpos=#2 \ypos=#3
\advance\ypos by\height
{\multiply\height by2
\puthmorphism(\xpos,\ypos)[#4`#5`{#7}]{\height}{\arrowtypea}a}%
\putmorphism(\xpos,\ypos)(1,-1)[`#6`{#8}]{\height}{\arrowtypeb}l%
\advance\xpos by\height
\advance\xpos by\height
\putmorphism(\xpos,\ypos)(-1,-1)[``{#9}]{\height}{\arrowtypec}r%
}}
 
\def\putVtriangle{\@ifnextchar <{\putVtrianglep}{\putVtrianglep
   <\arrowtypea`\arrowtypeb`\arrowtypec;\height>}}
\def\Vtriangle{\@ifnextchar <{\Vtrianglep}{\Vtrianglep
   <\arrowtypea`\arrowtypeb`\arrowtypec;\height>}}
\def\Vtrianglep<#1>[#2`#3`#4;#5`#6`#7]{{%%     %        #2------------->#3
\settriparms[#1]%                              %         \             /
\width=\height                                 %          \           /
\diagram%                                      %         #6\         /#7
\putVtrianglep<\arrowtypea`\arrowtypeb`%       %            \       /
\arrowtypec;\height>%                          %             \     /
(0,0)[#2`#3`#4;#5`#6`{#7}]%                    %              v   v
\enddiagram%%                                  %               #4
}}
 
\def\putVtrianglepairp<#1>(#2)[#3;#4`#5`#6`#7`#8]{{
\settripairparms[#1]%
\setpos(#2)%
\settokens`#3`%
\advance\ypos by\height
\putmorphism(\xpos,\ypos)(1,-1)[`\tokend`{#6}]{\height}{\arrowtypec}l%
\puthmorphism(\xpos,\ypos)[\tokena`\tokenb`{#4}]{\height}{\arrowtypea}a%
\advance\xpos by\height
\puthmorphism(\xpos,\ypos)[\phantom{\tokenb}`\tokenc`{#5}]%
{\height}{\arrowtypeb}a%
\putvmorphism(\xpos,\ypos)[``{#7}]{\height}{\arrowtyped}m%
\advance\xpos by\height
\putmorphism(\xpos,\ypos)(-1,-1)[``{#8}]{\height}{\arrowtypee}r%
}}
 
\def\putVtrianglepair{\@ifnextchar <{\putVtrianglepairp}{\putVtrianglepairp%
    <\arrowtypea`\arrowtypeb`\arrowtypec`\arrowtyped`\arrowtypee;\height>}}
\def\Vtrianglepair{\@ifnextchar <{\Vtrianglepairp}{\Vtrianglepairp%
    <\arrowtypea`\arrowtypeb`\arrowtypec`\arrowtyped`\arrowtypee;\height>}}
\def\Vtrianglepairp<#1>[#2;#3`#4`#5`#6`#7]{{%  %  #2a---->#2b---->#2c
\settripairparms[#1]%                          %   \      |      /
\settokens`#2`%                                %    \     |     /
\diagram%                                      %   #5\   #6    /#7
\putVtrianglepairp                             %      \   |   /
<\arrowtypea`\arrowtypeb`\arrowtypec`%         %       \  |  /
\arrowtyped`\arrowtypee;\height>%              %        v v v
(0,0)[{#2};#3`#4`#5`#6`{#7}]%                  %         #2d
\enddiagram%%
}}

\def\putCtrianglep<#1>(#2,#3)[#4`#5`#6;#7`#8`#9]{{%
\settriparms[#1]%
\xpos=#2 \ypos=#3
\advance\ypos by\height
\putmorphism(\xpos,\ypos)(1,-1)[``{#9}]{\height}{\arrowtypec}l%
\advance\xpos by\height
\advance\ypos by\height
\putmorphism(\xpos,\ypos)(-1,-1)[#4`#5`{#7}]{\height}{\arrowtypea}l%
{\multiply\height by 2
\putvmorphism(\xpos,\ypos)[`#6`{#8}]{\height}{\arrowtypeb}r}%
}}
 
\def\putCtriangle{\@ifnextchar <{\putCtrianglep}{\putCtrianglep
    <\arrowtypea`\arrowtypeb`\arrowtypec;\height>}}
\def\Ctriangle{\@ifnextchar <{\Ctrianglep}{\Ctrianglep
    <\arrowtypea`\arrowtypeb`\arrowtypec;\height>}}
\def\Ctrianglep<#1>[#2`#3`#4;#5`#6`#7]{{%%   %                / |
\settriparms[#1]%                            %             #5/  |
\width=\height                               %              /   |
\diagram%                                    %             v    |
\putCtrianglep<\arrowtypea`\arrowtypeb`%     %           #3     |#6
\arrowtypec;\height>%                        %             \    |
(0,0)[#2`#3`#4;#5`#6`{#7}]%                  %            #7\   |
\enddiagram%%                                %               \  |
}}                                           %                v v
\def\putDtrianglep<#1>(#2,#3)[#4`#5`#6;#7`#8`#9]{{%
\settriparms[#1]%
\xpos=#2 \ypos=#3
\advance\xpos by\height \advance\ypos by\height
\putmorphism(\xpos,\ypos)(-1,-1)[``{#9}]{\height}{\arrowtypec}r%
\advance\xpos by-\height \advance\ypos by\height
\putmorphism(\xpos,\ypos)(1,-1)[`#5`{#8}]{\height}{\arrowtypeb}r%
{\multiply\height by 2
\putvmorphism(\xpos,\ypos)[#4`#6`{#7}]{\height}{\arrowtypea}l}%
}}
 
\def\putDtriangle{\@ifnextchar <{\putDtrianglep}{\putDtrianglep
    <\arrowtypea`\arrowtypeb`\arrowtypec;\height>}}
\def\Dtriangle{\@ifnextchar <{\Dtrianglep}{\Dtrianglep
   <\arrowtypea`\arrowtypeb`\arrowtypec;\height>}}
\def\Dtrianglep<#1>[#2`#3`#4;#5`#6`#7]{{%%  %          | \
\settriparms[#1]%                           %          |  \#6
\width=\height                              %          |   \
\diagram%                                   %          |    v
\putDtrianglep<\arrowtypea`\arrowtypeb`%    %        #5|    #3
\arrowtypec;\height>%                       %          |    /
(0,0)[#2`#3`#4;#5`#6`{#7}]%                 %          |   /#7
\enddiagram%%                               %          |  /
}}                                          %          v v
\def\setrecparms[#1`#2]{\width=#1 \height=#2}%
\def\recursep<#1`#2>[#3;#4`#5`#6`#7`#8]{{\m@th
\width=#1 \height=#2
\settokens`#3`
\settowidth{\tempdimen}{$\tokena$}
\ifdim\tempdimen=0pt
  \savebox{\tempboxa}{\hbox{$\tokenb$}}%
  \savebox{\tempboxb}{\hbox{$\tokend$}}%
  \savebox{\tempboxc}{\hbox{$#6$}}%
\else
  \savebox{\tempboxa}{\hbox{$\hbox{$\tokena$}\times\hbox{$\tokenb$}$}}%
  \savebox{\tempboxb}{\hbox{$\hbox{$\tokena$}\times\hbox{$\tokend$}$}}%
  \savebox{\tempboxc}{\hbox{$\hbox{$\tokena$}\times\hbox{$#6$}$}}%
\fi
\ypos=\height
\divide\ypos by 2
\xpos=\ypos
\advance\xpos by \width
\bfig
\putCtrianglep<-1`1`1;\ypos>(0,0)[`\tokenc`;#5`#6`{#7}]%
\puthmorphism(\ypos,0)[\tokend`\usebox{\tempboxb}`{#8}]{\width}{-1}b%
\puthmorphism(\ypos,\height)[\tokenb`\usebox{\tempboxa}`{#4}]{\width}{-1}a%
\advance\ypos by \width
\putvmorphism(\ypos,\height)[``\usebox{\tempboxc}]{\height}1r%
\efig
}}
 
\def\recurse{\@ifnextchar <{\recursep}{\recursep<\width`\height>}}
 
\def\puttwohmorphisms(#1,#2)[#3`#4;#5`#6]#7#8#9{{%
\puthmorphism(#1,#2)[#3`#4`]{#7}0a
\ypos=#2
\advance\ypos by 20
\puthmorphism(#1,\ypos)[\phantom{#3}`\phantom{#4}`#5]{#7}{#8}a
\advance\ypos by -40
\puthmorphism(#1,\ypos)[\phantom{#3}`\phantom{#4}`#6]{#7}{#9}b
}}
 
\def\puttwovmorphisms(#1,#2)[#3`#4;#5`#6]#7#8#9{{%
\putvmorphism(#1,#2)[#3`#4`]{#7}0a
\xpos=#1
\advance\xpos by -20
\putvmorphism(\xpos,#2)[\phantom{#3}`\phantom{#4}`#5]{#7}{#8}l
\advance\xpos by 40
\putvmorphism(\xpos,#2)[\phantom{#3}`\phantom{#4}`#6]{#7}{#9}r
}}
 
\def\puthcoequalizer(#1)[#2`#3`#4;#5`#6`#7]#8#9{{%
\setpos(#1)%
\puttwohmorphisms(\xpos,\ypos)[#2`#3;#5`#6]{#8}11%
\advance\xpos by #8
\puthmorphism(\xpos,\ypos)[\phantom{#3}`#4`#7]{#8}1{#9}
}}
 
\def\putvcoequalizer(#1)[#2`#3`#4;#5`#6`#7]#8#9{{%
\setpos(#1)%
\puttwovmorphisms(\xpos,\ypos)[#2`#3;#5`#6]{#8}11%
\advance\ypos by -#8
\putvmorphism(\xpos,\ypos)[\phantom{#3}`#4`#7]{#8}1{#9}
}}
 
\def\putthreehmorphisms(#1)[#2`#3;#4`#5`#6]#7(#8)#9{{%
\setpos(#1) \settypes(#8)
\if a#9 %
     \vertsize{\tempcounta}{#5}%
     \vertsize{\tempcountb}{#6}%
     \ifnum \tempcounta<\tempcountb \tempcounta=\tempcountb \fi
\else
     \vertsize{\tempcounta}{#4}%
     \vertsize{\tempcountb}{#5}%
     \ifnum \tempcounta<\tempcountb \tempcounta=\tempcountb \fi
\fi
\advance \tempcounta by 60
\puthmorphism(\xpos,\ypos)[#2`#3`#5]{#7}{\arrowtypeb}{#9}
\advance\ypos by \tempcounta
\puthmorphism(\xpos,\ypos)[\phantom{#2}`\phantom{#3}`#4]{#7}{\arrowtypea}{#9}
\advance\ypos by -\tempcounta \advance\ypos by -\tempcounta
\puthmorphism(\xpos,\ypos)[\phantom{#2}`\phantom{#3}`#6]{#7}{\arrowtypec}{#9}
}}
 
\def\setarrowtoks[#1`#2`#3`#4`#5`#6]{%
\def\toka{#1}
\def\tokb{#2}
\def\tokc{#3}
\def\tokd{#4}
\def\toke{#5}
\def\tokf{#6}
}
\def\hex{\@ifnextchar <{\hexp}{\hexp<1000`400>}}
\def\hexp<#1`#2>[#3`#4`#5`#6`#7`#8;#9]{%
\setarrowtoks[#9]
\yext=#2 \advance \yext by #2
\xext=#1 \advance\xext by \yext
\bfig
\putCtriangle<-1`0`1;#2>(0,0)[`#5`;\tokb``\tokd]
\xext=#1 \yext=#2 \advance \yext by #2
\putsquare<1`0`0`1;\xext`\yext>(#2,0)[#3`#4`#7`#8;\toka```\tokf]
\advance \xext by #2
\putDtriangle<0`1`-1;#2>(\xext,0)[`#6`;`\tokc`\toke]
\efig
}
\makeatother

\let\ssection=\section
\renewcommand{\section}{\setcounter{equation}{0}\ssection}

\newtheorem{definition}{Definition}[section]

\newcommand\mapright[1]{\smash{
        \mathop{\mbox{\large{$\longrightarrow$}}}\limits^{#1}}}

\newcommand\mathC{\mkern1mu\raise2.2pt\hbox{$\scriptscriptstyle|$}
        {\mkern-7mu\rm C}}            % the complex numbers
\newcommand{\mathR}{{\rm I\! R}}      % the real numbers
\newcommand{\mathN}{{\rm I\!N}}       % the natural numbers

\newcommand\ie{{i.e.},}

\newcommand{\ga}{\gamma}
\newcommand{\Ga}{\Gamma}

\newcommand\s{\sigma}
\newcommand\Si{\Sigma}
\newcommand\De{\Delta}

\renewcommand{\O}{\Omega}
\newcommand{\id}{{\rm id}}
\newcommand\la{\langle}
\newcommand{\map}{\rightarrow}               % rightarrow (for maps)
\newcommand\ra{\rangle}

\newcommand{\op}{{\rm op}}              % op for opposite (category; in upright, not italic font)
\newcommand{\picl}{\pi_{{\rm cl}}}              % pi_cl, cl for classical (in upright, not italic font)
\renewcommand{\S}{{\cal S}}
\newcommand{\Hi}{{\cal H}}

\newcommand\BH{\mathcal{B(H)}}
\newcommand\PH{\mathcal{P(H)}}

\newcommand\typeTime{{\cal T}}
\newcommand\LeftDB{[\mkern-3mu[}
\newcommand\RightDB{]\mkern-3mu]}

\newcommand\eq[1]{(\ref{#1})}
\newcommand\eqs[2]{(\ref{#1}--\ref{#2})}
\newcommand\SAin[1]{\mbox{``}A\,\varepsilon\,#1\mbox{''}}
\newcommand\Ain[1]{A\,\varepsilon\,#1}
\newcommand\va[1]{\tilde{#1}}

\newcommand\dasBo[1]{\breve{\delta}^o(#1)}

\newcommand\q[1]{`#1\mbox{'}}
\renewcommand\L[1]{\mathcal{L}({#1})}
\newcommand\PL[1]{{\cal PL}(#1)}
\renewcommand\sp[1]{{\rm sp}(\hat A)}

\newcommand\Val[1]{\LeftDB\,#1\,\RightDB}
\newcommand\TVal[2]{\nu\big(#1;#2\big)}         % Truth value
\newcommand\Hom[3]{{\rm Hom}_{#1}\big(#2,#3\big)}

\newcommand\name[1]{\ulcorner #1\urcorner}      % name of sub-object
\newcommand\cha[1]{\chi_{#1}}                   % characteristic arrow

\newcommand\ps[1]{\underline{#1}}        % underline argument #1 (for presheaves)
\newcommand{\Om}{\ps{\Omega}}            % underlined Omega (presheaf of sieves, subobject classifier in cat of presheaves)
\newcommand{\dG}{\ps{\mkern1mu\raise2.5pt\hbox{$\scriptscriptstyle|$}
        {\mkern-7mu\rm O}}}                 % 'Coarse'de Groote presheaf
\newcommand{\dOU}{\ps{\mkern1mu\raise2.5pt\hbox{$\scriptscriptstyle|$}
        {\mkern-7mu\rm U}}}               % 'Coarse'de Groote presheaf

\newcommand{\Sig}{\ps{\Sigma}}            % Spectral presheaf
\newcommand{\R}{{\cal R}}                 % R (general quantity value object)

\newcommand{\SR}{\ps{{\mathR}^\succeq}}         % presheaf of real-valued order-reversing functions on V
\newcommand\F[1]{F_{\L{#1}}\big(\Sigma,\R\big)}
\newcommand\Ob[1]{{\rm Ob(#1)}}
\newcommand\Sub[1]{{\rm Sub}(#1)}              %subobjects of #1
\newcommand\fu[1]{\mathbf {#1}}

\newcommand\Set{{\bf Sets}}                    % Sets
\newcommand\SetH[1]{\Set^{{\V{#1}}^{\rm op}}}  % presheaf cat
\newcommand\SetC[1]{\Set^{{#1}^{\rm op}}}      % presheaf cat

\newcommand\V[1]{{\cal V}(\Hi_{#1})}           % V(H), the context cat

\begin{document}

\begin{titlepage}

\begin{center}
{\large\bf A Topos Foundation for Theories of Physics:

I. Formal Languages for Physics}
\end{center}

\begin{center}
        A.~D\"oring\footnote{email: a.doering@imperial.ac.uk}\\[10pt]

\begin{center}                      and
\end{center}

        C.J.~Isham\footnote{email: c.isham@imperial.ac.uk}\\[10pt]

        The Blackett Laboratory\\ Imperial College of Science,
        Technology \& Medicine\\ South Kensington\\ London SW7 2BZ\\
\end{center}

\begin{center}
      6 March 2007\end{center}

\begin{abstract}

This paper is the first in a series whose goal is to develop a
fundamentally new way of constructing theories of physics. The
motivation comes from a desire to address certain deep issues that
arise when contemplating quantum theories of space and time.

Our basic contention is that constructing a theory of physics is
equivalent to finding a representation in a topos of a certain
formal language that is attached to the system. Classical physics
arises when the topos is the category of sets. Other types of
theory employ a different topos.

In this paper we discuss two different types of language that can
be attached to a system, $S$. The first is a propositional
language, $\PL{S}$; the second is a higher-order, typed language
$\L{S}$.

Both languages provide deductive systems with an intuitionistic
logic. The reason for introducing $\PL{S}$ is that, as shown in
paper II of the series, it is the easiest way of understanding,
and expanding on, the earlier work on topos theory and quantum
physics. However, the main thrust of our programme utilises the
more powerful language $\L{S}$ and its representation in an
appropriate topos.
\end{abstract}
\end{titlepage}

\section{Introduction}
This paper is the first in a series whose goal is to develop a
fundamentally new way of constructing theories of physics. The
motivation comes from a desire to address certain deep issues that
arise when contemplating quantum theories of space and time.

A striking feature of the various current programmes for
quantising gravity---including superstring theory and loop quantum
gravity---is that, notwithstanding their disparate views on the
nature of space and time, they almost all use more-or-less
standard quantum theory. Although understandable from a pragmatic
viewpoint (since all we have \emph{is} more-or-less standard
quantum theory) this situation is nevertheless questionable when
viewed from a wider perspective. Indeed, there has always been a
school of thought asserting that quantum theory itself needs to be
radically changed/developed before it can be used in a fully
coherent quantum theory of gravity.

This iconoclastic stance has several roots, of which, for us, the
most important is the use in the standard quantum formalism of
certain critical mathematical ingredients that are taken for
granted and yet which, we claim, implicitly assume certain
properties of space and time. Such an \emph{a priori} imposition
of spatio-temporal concepts would be a major error  if they turn
out to be fundamentally incompatible with what is needed for a
theory of quantum gravity.

A prime example is the use of the \emph{continuum} which, in this
context, means  the real and/or complex numbers. These are a
central ingredient in all the various mathematical frameworks in
which quantum theory is commonly discussed. For example, this is
clearly so with the use of (i) Hilbert spaces and operators; (ii)
geometric quantisation; (iii) probability functions on a
non-distributive quantum logic; (iv) deformation quantisation; and
(v) formal (\ie\ mathematically ill-defined) path integrals and
the like. The \emph{a priori} imposition of such continuum
concepts could be radically incompatible with a quantum gravity
formalism in which, say, space-time is fundamentally discrete: as,
for example,  in  the causal set programme.

A secondary motivation for changing the quantum formalism is the
peristalithic problem of deciding how  a `quantum theory of
cosmology' could be interpreted  if one was lucky enough to find
one. Most people who worry about foundational issues in quantum
gravity would probably place the quantum cosmology/closed system
problem at, or near, the top of their list of reasons for
re-envisioning quantum theory. However, although we are certainly
interested in such conceptual issues,  the main motivation for our
research programme is \emph{not} to find a new interpretation of
quantum theory. Rather, the goal is to find a novel structural
framework within which new  \emph{types} of theory can  be
constructed, and in which continuum quantities play no fundamental
role.

Having said that, it is certainly true that the lack of any
external `observer'  of the universe `as a whole' renders
inappropriate the standard Copenhagen interpretation with its
instrumentalist use of counterfactual statements about what
\emph{would} happen \emph{if} a certain measurement was performed.
Indeed, the Copenhagen interpretation is inapplicable for
\emph{any}\footnote{Of course, the existence of the long-range,
and all penetrating, gravitational  force means that, at a
fundamental level, there is really only \emph{one} truly  closed
system, and that is the universe itself.} system that is truly
`closed' (or `self-contained') and for which, therefore, there is
no `external' domain in which an observer can lurk. This problem
has motivated much research over the years and continues to be of
wide interest. Clearly, the problem is particularly severe in a
quantum theory of cosmology.

When dealing with a closed system, what is needed is a
\emph{realist} interpretation of the theory, not one that is
instrumentalist.  The exact meaning of `realist' is infinitely
debatable but, when used by physicists, it typically means the
following:
\begin{enumerate}
\item The idea of `a property of the system' (\ie\ `the value of a
physical quantity') is meaningful, and representable  in the
theory.

\item Propositions about the system are handled using  Boolean
logic. This requirement is compelling in so far as we humans think
in a Boolean way.

\item There is  a space of `microstates' such  that specifying a
microstate\footnote{In simple non-relativistic systems, the state
is specified  at any given moment of time. Relativistic systems
(particularly quantum gravity!) require a more sophisticated
understanding of `state', but the general idea is the same.} leads
to unequivocal  truth values for all propositions  about the
system. The existence of such a state space is a natural way of
ensuring that the  first two requirements  are satisfied.
\end{enumerate}

The standard interpretation of classical physics satisfies these
requirements, and provides the paradigmatic example of a realist
philosophy in science. On the other hand,  the existence of  such
an interpretation in quantum theory is foiled by the famous
Kochen-Specker theorem \cite{KS67}.

What is needed is a formalism that is (i) free of \emph{prima
facie} prejudices about the nature of the values of  physical
quantities---in particular, there should be no fundamental use of
the real or complex numbers; and (ii) `realist', in at least  the
minimal  sense that propositions are meaningful, and  are assigned
`truth values', not just instrumentalist probabilities.

However, finding such a formalism is not easy: it is notoriously
difficult to modify the mathematical framework of quantum theory
without destroying the entire edifice. In particular, the Hilbert
space structure is very rigid and cannot easily be changed. And
the formal path-integral techniques do not fare much better.

Our approach includes finding a new way of formulating quantum
theory which, unlike the existing approaches, \emph{does} admit
radical generalisations and changes. A recent example of such an
attempt is the work of Abramsky and Coecke who construct a
categorical analogue of some of the critical parts of the Hilbert
space formalism \cite{AC04}; see also the work by Vicary
\cite{Vic06}. Here, we adopt a different strategy based on the
intrinsic logical structure that is associated with any
topos.\footnote{Topos theory is a sophisticated subject and, for
theoretical physicists, not always that easy to understand. The
references that we have found most helpful in this series of
papers are \cite{McL71,Gol84,LamScott86,Bell88,MM92,Jst02}. Some
of the basic ideas are described briefly in the Appendix to this
paper.}

Our  contention is that theories of a  physical system should be
formulated in a topos that depends on both the theory-type  and
the system. More precisely, if  a theory-type (such as classical
physics, or quantum physics) is applicable to a certain class of
systems,  then, for each system in this class, there is a topos in
which  the theory is to be formulated. For some theory-types the
topos is system-independent: for example, conventional classical
physics always uses the topos of sets. For other theory-types, the
topos varies from system to system: for example, this is the case
in quantum theory.

In regard to the three conditions listed above for a `realist'
interpretation, our scheme has the following ingredients:
\begin{enumerate}
\item The concept of the `value of a physical quantity' is
meaningful, although this `value' is associated with an object in
the topos that may not be the real-number object. With that
caveat, the concept of a `property of the system' is also
meaningful.

\item Propositions about a system are representable by a Heyting
algebra associated with the topos. A Heyting algebra is a
distributive lattice that differs from a Boolean algebra only in
so far as the \emph{law of excluded middle} need not hold, \ie\ $
\alpha\lor\lnot \alpha\preceq1$. A Boolean algebra is a Heyting
algebra with strict equality: $\alpha\lor\lnot\alpha=1$.

\item There is a `state object' in the topos. However,  generally
speaking, there will not be enough `microstates' to determine
this. Nevertheless, truth values can be assigned to propositions
with the aid of a `truth object'.  These truth values lie in
another Heyting algebra.
\end{enumerate}

This new approach affords a way in which it becomes feasible to
generalise quantum theory  without any fundamental reference to
Hilbert spaces, path integrals, etc.; in particular, there is no
\emph{prima facie} reason for introducing continuum quantities. As
we have emphasised, this is our main motivation for developing the
topos approach. We shall say more about this later.

From a conceptual perspective, a central feature of our scheme is
the `neo-realist\footnote{We coin the term `neo-realist' to
signify the conceptual structure implied by our topos formulation
of theories of physics.}' structure reflected in the three
statements above. This neo-realism is the conceptual fruit of  the
mathematical fact that a physical theory  expressed in a topos
`looks' very  much like \emph{classical} physics.

This fundamental feature stems from (and, indeed, is defined by)
the existence of two special objects in the topos:  the `state
object'\footnote{The meaning of the subscript `$\phi$' is
explained in the main text. It refers to a particular
topos-representation of a formal language attached to the system:
see later.}, $\Si_\phi$, mentioned above, and the `quantity-value
object', $\R_\phi$. Then: (i) any physical quantity, $A$, is
represented by an arrow $A_\phi:\Si_\phi\map\R_\phi$ in the topos;
and (ii) propositions about the system are represented by
sub-objects of the state object $\Si_\phi$. These form a Heyting
algebra, as is the case for the set of sub-objects of any object
in a topos.

The fact that physical quantities are  represented by arrows whose
domain is the object $\Si_\phi$, and propositions are represented
by sub-objects of $\Si_\phi$, suggests strongly that $\Si_\phi$
should be regarded as the topos-analogue of a classical  state
space. Indeed, for any classical system the topos is just the
category of sets, $\Set$, and then the ideas above  reduce to the
familiar picture in which (i) there is a state space $\S$ which is
a set; (ii) any physical quantity, $A$,  is represented by a
real-valued functions $\breve{A}:\S\map\mathR$; and (iii)
propositions are represented by subsets of $\S$, and with the
associated Boolean algebra.

The present work is the first of a series of papers  devoted to
exploring in depth the idea that theories of physics should be
expressed in a topos that depends on both the theory-type and the
system; and that physical quantities and propositions are
represented in the ways indicated above.  Papers II  and III in
the series  are concerned with  quantum theory \cite{DI(2),DI(3)}
which serves as a paradigmatic example for the general theory.
These ideas are motivated by earlier work by one of us (CJI) and
Butterfield on interpreting quantum theory in a topos \cite{IB98,
IB99, IB00, IB02, Ish05, IB00b}; see also \cite{Isham97,Doe05}.

In the present paper,  we will make precise the sense in which
propositions about a system can be represented by sub-objects of
an object in a topos. To this end, we introduce  a formal language
for each system with the key idea that the construction of a
theory of the system involves finding a representation of the
associated language in an appropriate topos. These languages are
deductive systems employing intuitionistic logic; as  such, they
can be used to make, and manipulate, statements about the world as
it is revealed in the system under study.

In paper IV (\cite{DI(4)}) we return once more to the overall
formalism and consider what happens to the languages and their
representations when the system   ranges over the objects in a
`category of systems'. This category incorporates the ideas of
forming composites of systems, and finding sub-systems of a
system.

The plan of the present paper is as follows. Section
\ref{Sec:ConceptualBackground} is written in a rather discursive
style and deals with various topics with a significant conceptual
content. In particular, we discuss in more detail some of the
issues concerning the status of continuum quantities in physics.

Then, in Section \ref{Sec:ToposLogic} we introduce a simple
propositional language, $\PL{S}$, that can be used to assert
statements about the world as it is reflected in the system $S$.
The propositional logic used in this language is intuitionistic
and, therefore, it is mathematically consistent to seek
representations  of $\PL{S}$ in a Heyting algebra; in particular
in the collection of sub-objects of the state object of a topos.

Simple propositional languages are limited in scope and,
therefore,  in Section \ref{Sec:TypedLanguage} a higher-order,
typed language, $\L{S}$, is developed. Languages of this sort lie
at the heart of topos theory and are of great power. We discuss in
detail an example of such a language which, although simple, can
be used for many physical systems. This language has just two
`ground type' symbols, $\Si$ and $\R$, that are the linguistic
precursors of the state object, and quantity-value object,
respectively. In addition, there are `function symbols'
$A:\Si\map\R$ that represent physical quantities in the theory. We
show how representations of $\L{S}$ in a topos correspond to
concrete physical theories, and work  out the scheme in detail for
classical physics. (The application to quantum theory is discussed
in the next two papers \cite{DI(2), DI(3)}.) Finally, in Section
\ref{Sec:Concl} we draw some conclusions about this first chapter
of our endeavour to construct a topos framework within which to
construct theories of physics.

The paper concludes with an Appendix which contains some of the
central ideas of topos theory. Many important topics are left out
for reasons of space, but we have tried to include the key ideas
used in this series of papers. To gain a proper understanding of
topos theory, we recommend the standard text books
\cite{McL71,Gol84,LamScott86,Bell88,MM92,Jst02}

\section{The Conceptual Background of our Scheme}
\label{Sec:ConceptualBackground}
\subsection{The Problem of Using Real Numbers a Priori}
As mentioned in the Introduction, one of the main goals of our
work is to find new tools with which to develop theories that are
significant extensions of, or developments from, quantum theory
but without being tied \emph{a priori} to the use of the real or
complex numbers.

In this context we note  that  real numbers  arise in theories of
physics in three different (but related) ways: (i)  as the values
of physical quantities; (ii) as the values of probabilities; and
(iii) as a fundamental ingredient in models of space and time
(especially in those based on differential geometry). The first
two are of direct concern in our worries about making unjustified,
\emph{a priori} assumptions in quantum theory, and we shall now
examine them in detail.

\paragraph{Why are physical quantities assumed to be real-valued?}
One reason for assuming physical quantities to be real-valued is
undoubtedly  that, traditionally (\ie\ in the pre-digital age),
they are measured with rulers and pointers (or they are defined
operationally in terms of such measurements), and rulers and
pointers are taken to be classical objects that exist in the
continuum physical space of classical physics. In this sense there
is a direct link between the space in which physical quantities
take their values   (what we shall call the `quantity-value
space') and the nature of physical space or space-time
\cite{Isham03}.

If  conceded, this claim means that the assumption that physical
quantities are real-valued is problematic in a theory in which
space, or space-time, is not modelled by a smooth manifold.
Admittedly, if the theory employs a \emph{background} space, or
space-time---and if this background is a manifold---then the use
of real-valued physical quantities \emph{is} justified in so far
as their value-space can be related to this background. Such a
stance is particularly appropriate in situations where the
background plays a central role in giving meaning to concepts like
`observers' and  `measuring devices', and thereby provides a basis
for an instrumentalist interpretation of the theory.

However, caution is needed with this argument since the background
structure may arise only in some `sector' of the theory; or  it
may exist only in some limiting, or approximate, sense. The
associated instrumentalist interpretation would then be similarly
limited in scope.  For this reason, if no other, a `realist'
interpretation is more attractive than an instrumentalist one.

In fact, in  such circumstances, the phrase `realist
interpretation' does not really do justice to the situation since
it tends to imply that there are other interpretations of the
theory, particularly instrumentalism, with which the realist one
can contend on a more-or-less equal footing. But, as we just
argued, the instrumentalist interpretation may be severely limited
in scope as compared   to the realist one. To flag this point,  we
will sometimes refer to a `realist formalism', rather than a
`realist interpretation'.\footnote{Of course, such discussions are
unnecessary in classical physics since, there, if knowledge of the
value of a physical quantity is gained by making a (ideal)
measurement, the reason why we obtain the result that we do, is
because the quantity \emph{possessed} that value immediately
before the measurement was made. In other words, ``epistemology
models ontology''---a slogan employed with great enthusiasm by
John Polkinghorne in his advocacy of the philosophy of `critical
realism' as a crucial tool with which to analyse epistemological
parallels between science and religion. Supposedly, the phrase is
printed on his T-shirts:-)}

\paragraph{Why are probabilities required to lie in the
interval $[0,1]$?}  The motivation for using  the subset $[0,1]$
of the real numbers as the value space for probabilities comes
from the relative-frequency interpretation of probability. Thus,
in principle, an experiment is to be repeated a large number, $N$,
times, and the probability associated with a particular result is
defined to be the ratio $N_i/N$, where $N_i$ is the number of
experiments in which that result was obtained. The rational
numbers $N_i/N$ necessarily lie between $0$ and $1$, and if the
limit $N\map\infty$ is taken---as is appropriate for a
hypothetical `infinite ensemble'---real numbers in the closed
interval $[0,1]$ are obtained.

The relative-frequency interpretation of probability is natural in
instrumentalist theories of physics,  but it is not meaningful if
there is no classical spatio-temporal background in which the
necessary  measurements could be made; or, if there is  a
background, it is one to which the relative-frequency
interpretation cannot be adapted.

In the absence of   a relativity-frequency interpretation, the
concept of `probability' must be understood in a different way. In
the physical sciences, one of the most discussed approaches
involves  the concept of `potentiality', or `latency', as favoured
by Heisenberg, Margenau, and Popper
\cite{Heisenberg52}\cite{Margenau49}\cite{Popper82} (and, for good
measure, Aristotle). In this case there is no  compelling reason
why the probability-value space should be a subset of the real
numbers. The minimal requirement is that this value-space is an
ordered set---so that  one proposition can be said to be more or
less probable than another. However, there is no \emph{prima
facie} reason why this set should be \emph{totally} ordered: \ie\
there may be  pairs of propositions whose potentialities  cannot
be compared---something that seems eminently plausible in the
context of non-commensurable quantities in quantum theory.

By invoking the idea of `potentiality', it becomes feasible to
imagine a quantum-gravity theory with no spatio-temporal
background but where probability is still a fundamental concept.
However,  it could also be that the  concept of probability plays
no fundamental role in such circumstances, and can be given a
meaning \emph{only} in the context of a sector, or limit, of the
theory where a background does exist. This background could then
support a limited instrumentalist interpretation which  would
include  a (limited) relative-frequency understanding of
probability.

In fact, most modern approaches  to quantum gravity aspire to  a
formalism that is background independent
\cite{Bae00,CM05,Smo05,Smo06}. So, if a background space  does
arise, it will  be in one of the restricted senses mentioned
above.  Indeed, it is often asserted that a proper theory of
quantum gravity will not involve \emph{any} direct spatio-temporal
concepts, and that  what we commonly call `space' and `time' will
`emerge' from the formalism only in some appropriate limit
\cite{BI01}. In  this case, any instrumentalist interpretation
could only `emerge' in the same limit, as would the associated
relative-frequency interpretation of probability.

In a theory of this type, there will be no \emph{prima facie} link
between the values of physical quantities and the nature of space
or space-time although, of course, this cannot be totally ruled
out. In any event, part of the fundamental specification of the
theory will involve deciding what the `quantity-value space'
should be.

These considerations suggest that quantum theory must be radically
changed in order to  accommodate situations where there is no
background  space, or space-time, manifold within which an
instrumentalist interpretation can be formulated, and where,
therefore, some sort of `realist' formalism is essential.

These reflections also suggest  that the quantity-value space
employed in  an instrumentalist realisation of a theory---or a
`sector', or `limit', of the theory---need not be the same as the
quantity-value space in a neo-realist formulation. At first sight
this may seem strange but, as is shown in the third paper of this
series, this is precisely what happens in the topos reformulation
of standard quantum theory \cite{DI(3)}.

\subsection{The Genesis of Topos Ideas in Physics}
\subsubsection{A Possible Role for Heyting Algebras}
To motivate  topos theory as the source of neo-realism let us
first consider classical physics, where everything is defined in
the category, $\Set$, of sets and functions between sets. Then (i)
any physical quantity, $A$, is represented by a real-valued
function $\breve{A}:\S\map\mathR$, where $\S$ is the space of
microstates; and (ii) a proposition of the form $\SAin\De$ (which
asserts that the value of the physical quantity $A$ lies in the
subset $\De$ of the real line $\mathR$)\footnote{In the rigorous
theory of classical physics, the set $\S$ is a symplectic
manifold, and $\De$ is a \emph{Borel} subset of $\mathR$. Also,
the function
 $\breve{A}:{\cal S}\map\mathR$  may be required to be
measurable, or continuous, or smooth, depending on the quantity,
$A$, under consideration.} is represented by the
subset\footnote{Throughout this series of papers we will adopt the
notation in which $A\subseteq B$ means that $A$ is a subset of $B$
that could equal $B$; while $A\subset B$ means that $A$ is a
\emph{proper} subset of $B$; \ie\ $A$ does not equal $B$. Similar
remarks apply to other pairs of ordering symbols like
$\prec,\preceq$; or $\succ,\succeq$, etc.}
$\breve{A}^{-1}(\De)\subseteq \S$. In fact any proposition $P$
about the system is represented by an associated subset, $\S_P$,
of $\S$: namely, the set of states for which $P$ is true.
Conversely, every subset of $\S$ represents a
proposition.\footnote{More precisely, every Borel subset of $\S$
represents \emph{many} propositions about the values of physical
quantities. Two propositions are said to be `physically
equivalent' if they are represented by the same subset of $\S$.}

It is easy to see how the logical calculus of propositions arises
in this picture. For let $P$ and $Q$ be propositions, represented
by the subsets $\S_P$ and $\S_Q$ respectively, and consider the
proposition ``$P$ and $Q$''. This is true if, and only if, both
$P$ and $Q$ are true, and hence the subset of states that
represents this logical conjunction consists of those states that
lie in both $\S_P$ and $\S_Q$---\ie\ the set-theoretic
intersection $\S_P\cap\S_Q$. Thus ``$P$ and $Q$'' is represented
by $\S_P\cap\S_Q$. Similarly, the proposition ``$P$
 or $Q$'' is true if either $P$ or $Q$ (or both) are true,
and hence this logical disjunction is represented by those states
that lie in $\S_P$ plus those states that lie in $\S_Q$---\ie\ the
set-theoretic union $\S_P\cup\S_Q$. Finally, the logical negation
``not $P$'' is represented by all those points in $\S$ that do not
lie in $\S_P$---\ie\ the set-theoretic complement $\S/\S_P$.

In this way, a fundamental relation is established between the
logical calculus of propositions about a physical system, and the
Boolean algebra of subsets of the state space. Thus the
mathematical structure of classical physics is such that, \emph{of
necessity}, it reflects a `realist' philosophy, in the sense in
which we are using the word.

One way to escape from the tyranny of Boolean algebras and
classical realism is via topos theory. Broadly speaking, a topos
is a category that behaves very much like the category of sets
(see Appendix); in particular, the collection of sub-objects of an
object forms a \emph{Heyting algebra}, just as the collection of
subsets of a set form a Boolean algebra. Our intention, therefore,
is to explore the possibility of associating physical propositions
with sub-objects of some object $\Si$ (the analogue of a classical
state space) in some topos.

A Heyting algebra, $\mathfrak{h}$, is  a distributive lattice with
a zero element, $0$, and a unit element, $1$, and with the
property that to each pair $\alpha,\beta\in\mathfrak{h}$ there is
an implication $\alpha\Rightarrow\beta$, characterized by
\begin{equation}
\ga\preceq(\alpha\Rightarrow\beta)\mbox{ if and only if } \ga\land
\alpha\preceq \beta.
\end{equation}
The negation is defined as $\lnot\alpha:=(\alpha\Rightarrow0)$ and
has the property that the \emph{law of excluded middle} need not
hold, \ie\ there may exist $\alpha\in\mathfrak{h}$, such that $
\alpha\lor\lnot \alpha\prec1$ or, equivalently, $\lnot\lnot
\alpha\succ\alpha$. This is the characteristic property of an
intuitionistic logic. A Boolean algebra is the special case of a
Heyting algebra in which there is the strict equality
$\alpha\lor\lnot\alpha=1$.

The elements of a Heyting algebra can be manipulated in a very
similar way to those in a Boolean algebra. One of our claims is
that,  as far as theories of physics are concerned, Heyting logic
is a  viable\footnote{The main difference between theorems proved
using Heyting logic and those using Boolean logic is that proofs
by contradiction cannot be used in the former. In particular, this
means that one cannot prove that something exists by arguing that
the assumption that it does not  leads to contradiction; instead
it is necessary to provide a \emph{constructive} proof of the
existence of the entity concerned. Arguably, this does not place
any major restriction on building theories of physics.}
alternative to Boolean logic.

To give some idea of the difference between a Boolean algebra and
a Heyting algebra, we note that the paradigmatic example of the
former is the collection of all measurable subsets of a measure
space $X$. Here, if $\alpha\subseteq X$ represents a proposition,
the logical negation, $\neg\alpha $, is just the set-theoretic
complement $X/\alpha$.

On the other hand, the paradigmatic example of a Heyting  algebra
is the collection of all open sets in a topological space $X$.
Here, if $\alpha\subseteq X$ is open, the logical negation
$\neg\alpha$ is defined to be the \emph{interior} of the
set-theoretical complement $X/\alpha$. Therefore, the difference
between $\neg\alpha$ in the topological space $X$, and
$\lnot\alpha$ in the measurable space generated by the topology of
$X$, is just the `thin' boundary of $X/\alpha$.

\subsubsection{Our Main Contention about Topos Theory and Physics}
We contend that, for a given theory-type (for example, classical
physics, or quantum physics), each system $S$ to which the theory
is applicable is associated with a particular topos $\tau_\phi(S)$
within whose framework the theory, as applied to $S$, is to be
formulated and interpreted. In this context, the
`$\phi$'-subscript is a label that changes as the theory-type
changes. It signifies the representation of a system-language in
the topos $\tau_\phi(S)$: we will come to this later.

The conceptual interpretation of this formalism is `neo-realist'
in the following sense:
\begin{enumerate}
\item A physical quantity, $A$, is represented by an arrow
$A_{\phi,S}:\Si_{\phi,S}\map\R_{\phi,S}$ where $\Si_{\phi,S}$ and
$\R_{\phi,S}$ are two special objects in the topos $\tau_\phi(S)$.
These are the analogues of, respectively, (i) the classical state
space, $\cal S$; and (ii) the real numbers, $\mathR$, in which
 classical physical quantities take their values.

In what follows,  $\Si_{\phi,S}$ and $\R_{\phi,S}$ are called the
`state object', and the `quantity-value object', respectively.

\item Propositions about the system $S$ are represented by
sub-objects of $\Si_{\phi,S}$. These sub-objects form a Heyting
algebra.

\item Once the topos analogue of a state (a `truth object') has
been specified,  these propositions are assigned truth values in
the Heyting logic associated with the global elements of the
sub-object classifier, $\O_{\tau_\phi(S)}$, in the topos
$\tau_\phi(S)$.
\end{enumerate}

Thus a theory expressed in this way \emph{looks} very much like
classical physics except that whereas classical physics always
employs the topos of sets, other theories---including quantum
theory and, we conjecture, quantum gravity---use a different
topos.

One deep result in topos theory is that there is an \emph{internal
language} associated with each topos. In fact, not only does each
topos generate an internal language, but, conversely, a language
satisfying appropriate conditions generates a topos. Topoi
constructed in this way are called `linguistic topoi', and every
topos can be regarded as a linguistic topos. In many respects,
this is one of the profoundest  ways of understanding what a topos
really `is'. This aspect of topos theory is discussed at length in
the books by Bell \cite{Bell88}, and Lambek and Scott
\cite{LamScott86}.

These results are exploited  in Section \ref{Sec:TypedLanguage}
where we introduce the idea that, for any applicable\ theory-type,
each physical system $S$  is associated with a `local' language,
$\L{S}$. The application of the theory-type to $S$ is then
equivalent to finding a \emph{representation} of $\L{S}$ in a
topos.

Closely related to the existence of this linguistic structure is
the striking fact that a topos can be used as a \emph{foundation}
for  mathematics itself, just as set theory is used in the
foundations of `normal' (or `classical') mathematics. In this
context, the key remark  is that the internal language of a topos
has a form that is similar in many ways to the formal language on
which normal set theory is based. It is this internal, topos
language that is used to \emph{interpret} the theory in a
`neo-realist' way.

The main difference with classical logic is that the logic of the
topos language does not satisfy the principle of excluded middle,
and  hence proofs by contradiction are not permitted. This has
many intriguing consequences. For example, there are topoi with
genuine \emph{infinitesimals} that can be used to construct a
rival to normal calculus. The possibility of such quantities stems
from the fact that the normal proof that they do \emph{not} exist
is a proof by contradiction.

Thus each topos carries its own world of mathematics: a world
which, generally speaking, is \emph{not} the same as that of
classical mathematics.

Consequently, by postulating that, for a given theory-type, each
physical system carries its own topos, we are also saying that to
each physical system plus theory-type there is associated a
framework for mathematics itself! Thus classical physics uses
classical mathematics; and quantum theory uses `quantum
mathematics'---the mathematics formulated in the topoi of quantum
theory. To this we might add the conjecture: ``Quantum gravity
uses `quantum gravity' mathematics''!

\section{Propositional Languages and Theories of Physics}
\label{Sec:ToposLogic}
\subsection{Two Opposing Interpretations of Propositions}
\label{SubSec:BackgroundRemarks} Attempts to construct a na\"ive
realist interpretation of quantum theory founder on the
Kochen-Specker theorem. However, if, despite this theorem, some
degree of realism is still sought, there are not that many
options.

One approach is to `reify' only a subset of physical variables,
as, for example, in the pilot-wave approach and other `modal
interpretations'. A topos-theoretic extension of this idea of
`partial reification' was   proposed in \cite{IB98, IB99, IB00,
IB02} with a technique in which all possible reifyable sets of
physical variables are included on an equal footing. This involves
constructing a category, $\cal C$, whose objects are collections
of quantum observables that \emph{can} be simultaneously reified
because the corresponding self-adjoint operators commute. Of
course, the concept of `measurement' plays no fundamental role in
our neo-realist, topos approach.

In this earlier work, it was postulated that the logic for
handling quantum propositions from this perspective is that
associated with the topos of presheaves\footnote{The idea of a
presheaf is discussed briefly in the Appendix. From a physical
perspective, the objects in the category $\cal C$ are
\emph{contexts} in which the structure of the theory can be
discussed. In quantum theory,  the category $\cal C$ is just a
partially-ordered set, which simplifies many manipulations.},
$\Set^{{\cal C}^\op}$. The idea is that a single presheaf will
encode any quantum proposition from the perspective of \emph{all}
contexts at once. However, in the original papers, the crucial
`daseinisation' operation (see paper II) was not known and,
consequently, the discussion became convoluted in places. In
addition, the generality and power of the underlying procedure was
not fully appreciated.

For this reason, in the present paper we   return to the basic
questions and reconsider them in the light of the overall topos
structure that has now become clear.

We start by considering the way in which  propositions arise, and
are manipulated, in physics. For simplicity, we will concentrate
on systems that are associated with `standard' physics. Then, to
each such system $S$ there is associated a set of physical
quantities---such as energy, momentum, position, angular momentum
etc.\footnote{This set does not have to contain `\emph{all}'
possible physical quantities: it suffices to concentrate on a
subset that are deemed to be of particular interest. However, at
some point, questions may arise about the `completeness' of  the
set.}---all of which are real-valued.   The associated
propositions are of the form $\SAin\De$, where $A$ is a physical
quantity, and $\De$ is a subset\footnote{For various reasons, the
subset $\De\subseteq\mathR$ is usually required to be a
\emph{Borel} subset, and we will assume this without further
comment.} of $\mathR$.

From a conceptual perspective, the proposition $\SAin\De$ can be
read in two, very different, ways:
\begin{enumerate}
\item[(i)] {\bf The (na\"ive) realist interpretation:} ``The
physical quantity $A$ \emph{has} a value, and that value lies in
$\De$.''

\item[(ii)] {\bf The instrumentalist interpretation:}  ``\emph{If}
a measurement is made of $A$, the result will be found to lie in
$\De$.''
\end{enumerate}
The former is the familiar, `commonsense' understanding of
propositions in both classical physics and daily life. The latter
underpins the Copenhagen interpretation of quantum theory. The
instrumentalist interpretation can, of course, be applied to
classical physics too, but it does not lead to anything new. For,
in classical physics, what is measured is what \emph{is} the case:
``Epistemology models ontology''.

We will now study the role of propositions in physics more
carefully, particularly in the context of  `realist'
interpretations.

\subsection{The Propositional Language $\PL{S}$}
\label{SubSec:PropLangPhys}
\subsubsection{Intuitionistic Logic and the Definition of $\PL{S}$}
We are going to construct a formal language, $\PL{S}$, with which
to express propositions about a physical system, $S$, and to make
deductions concerning them. Our intention is to interpret these
propositions  in a `realist' way: an endeavour whose  mathematical
underpinning  lies in constructing a representation of  $\PL{S}$
in a Heyting algebra, $\mathfrak{H}$, that is part of the
mathematical framework involved in the application of a particular
theory-type to $S$.

The first step is to construct the set, $\PL{S}_0$,  of all
strings of the form $\SAin\De$ where $A$ is a physical quantity of
the system $S$, and $\De$ is a (Borel) subset of the real line,
$\mathR$. Note that what has here been called a `physical
quantity' could better (but more clumsily) be termed the `name' of
the physical quantity. For example, when we talk about the
`energy' of a system, the word `energy' is the same, and functions
in the same way in the formal language, irrespective of the
details of the actual Hamiltonian of the system.

The strings $\SAin\De$ are taken to be the \emph{primitive
propositions} about the system,  and are used  to define
`sentences'. More precisely, a new set of symbols
$\{\neg,\land,\lor,\Rightarrow\}$ is added to the language, and
then a \emph{sentence} is defined inductively by the following
rules (see Ch. 6 in \cite{Gol84}):
\begin{enumerate}
\item Each primitive proposition $\SAin\De$ in $\PL{S}_0$ is a
sentence.

\item If $\alpha$ is a sentence, then so is $\neg\alpha$.

\item If $\alpha$ and $\beta$ are sentences, then so are
$\alpha\land\beta$, $\alpha\lor\beta$, and
$\alpha\Rightarrow\beta$.
\end{enumerate}
The collection of all sentences, $\PL{S}$, is an elementary formal
language that can be used to express and manipulate propositions
about the system $S$.  Note that the symbols $\neg$, $\land$,
$\lor$, and $\Rightarrow$ have no explicit meaning, although of
course the implicit intention is that they should stand for `not',
`and', `or' and `implies', respectively. This implicit meaning
becomes explicit when a representation of $\PL{S}$ is constructed
as part of the application of a theory-type to $S$ (see below).
Note also that $\PL{S}$ is a \emph{propositional} language only:
it does not contain the quantifiers `$\forall$' or `$\exists$'. To
include them requires a higher-order language. We shall return to
this  in our discussion of the local language $\L{S}$.

The next step arises because $\PL{S}$ is not only a vehicle for
expressing propositions about the system $S$: we also want to
\emph{reason} with it about the system. To achieve this, a series
of axioms for a deductive logic  must be added to $\PL{S}$. This
could be either classical logic or intuitionistic logic, but we
select the latter since it allows a larger class of
representations/models, including representations in topoi in
which the law of excluded middle fails.

The axioms for intuitionistic logic consist of a finite collection
of sentences in $\PL{S}$ (for example,
$\alpha\land\beta\Rightarrow\beta\land\alpha$), plus  a single
rule of inference, \emph{modus ponens} (the `rule of detachment')
which says that from $\alpha$ and $\alpha\Rightarrow\beta$ the
sentence $\beta$ may be derived.

Others axioms  might be added to $\PL{S}$ to reflect the implicit
meaning of  the primitive proposition $\SAin\De$: \ie\ ``$A$ has a
value, and that value lies in $\De\subseteq\mathR$''. For example,
the sentence ``$\Ain{\De_1} \land \Ain{\De_2}$'' (`$A$ belongs to
$\De_1$' \emph{and} `$A$ belongs to $\De_2$')  might seem to be
equivalent to ``$\Ain{\De_1\cap\De_2}$'' (`$A$ belongs to
$\De_1\cap\De_2$'). A similar remark applies to ``$\Ain{\De_1}\lor
\Ain{\De_2}$''.

Thus, along with the axioms of intuitionistic logic and
detachment, we might be tempted to add the following axioms:
\begin{eqnarray}
\Ain{\De_1}\land \Ain{\De_2}&\Leftrightarrow&
                     \Ain{\De_1\cap\De_2} \label{AinD1andD2}\\
\Ain{\De_1}\lor \Ain{\De_2}&\Leftrightarrow&
                     \Ain{\De_1\cup\De_2} \label{AinD1orD2}
\end{eqnarray}
These axioms are consistent with the intuitionistic logical
structure of $\PL{S}$.

We shall see later the extent to which the axioms
\eqs{AinD1andD2}{AinD1orD2} are compatible with the topos
representations of classical physics, and of quantum physics.
However, the other obvious proposition to consider in this
way---``It is \emph{not} the case that $A$ belongs to $\De$''---is
clearly problematical.

In classical logic, this proposition\footnote{The parentheses
$(\;)$ are not  symbols in the language; they are just a way of
grouping letters and sentences.}, ``$\neg(\Ain\De)$'', is
equivalent to ``$A$ belongs to $\mathR\backslash\De$'', where
$\mathR\backslash\De$ denotes the set-theoretic complement of
$\De$ in $\mathR$. This  suggests augmenting
\eqs{AinD1andD2}{AinD1orD2} with a third axiom
\begin{equation}
    \neg(\Ain\De) \Leftrightarrow \Ain{\mathR\backslash\De}
                                                \label{negAinD}
\end{equation}
However, applying `$\neg$' to both sides of \eq{negAinD} gives
\begin{equation}
        \neg\neg(\Ain\De) \Leftrightarrow \Ain\mathR
\end{equation}
because of the set-theoretic result
$\mathR\backslash(\mathR\backslash\De)=\De$. But in an
intuitionistic logic we do not have
$\alpha\Leftrightarrow\neg\neg\alpha$ but only
$\alpha\Rightarrow\neg\neg\alpha$, and so \eq{negAinD} could be
false in a Heyting-algebra representation of $\PL{S}$ that was not
Boolean. Therefore, adding \eq{negAinD} as an axiom in $\PL{S}$ is
not indicated if representations are  to be sought in non-Boolean
topoi.

\subsubsection{Representations of $\PL{S}$.}
To use the language $\PL{S}$  `for real' it must be represented in
the concrete mathematical structure that arises when a theory-type
is applied to $S$. Such a representation, $\pi$, maps each of the
primitive propositions, $\alpha$, in $\PL{S}_0$ to an element,
$\pi(\alpha),$ of some Heyting algebra (which could be Boolean),
$\mathfrak{H}$, whose specification is, of course, part of the
theory. For example, in classical mechanics, the propositions are
represented in the Boolean algebra of all (Borel) subsets of the
classical state space.

The representation of the primitive propositions can be extended
recursively to all of $\PL{S}$ with the aid of the following rules
\cite{Gol84}:
\begin{eqnarray}
&(a)& \pi(\alpha\lor\beta):=\pi(\alpha)\lor\pi(\beta)
                                                \label{pi(a)}\\
&(b)& \pi(\alpha\land\beta):=\pi(\alpha)\land\pi(\beta)
                                                \label{pi(b)}\\
&(c)& \pi(\neg\alpha):=\neg\pi(\alpha)\hspace{3cm} \label{pi(c)} \\
&(d)&
\pi(\alpha\Rightarrow\beta):=\pi(\alpha)\Rightarrow\pi(\beta)
                                                \label{pi(d)}
\end{eqnarray}
Note that,  on the left hand side of \eqs{pi(a)}{pi(d)}, the
symbols $\{\neg,\land,\lor,\Rightarrow\}$ are elements of the
language $\PL{S}$, whereas on the right hand side they are the
logical connectives in the Heyting algebra, ${\cal \mathfrak{H}}$,
in which the representation takes place.

This extension of $\pi$ from $\PL{S}_0$ to $\PL{S}$ is consistent
with the axioms for the intuitionistic, propositional logic  of
the language $\PL{S}$. More precisely, these axioms become
tautologies: \ie\ they are all represented by the maximum element,
$1$, in the Heyting algebra. By construction, the map
$\pi:\PL{S}\map\cal \mathfrak{H}$ is then a representation of
$\PL{S}$ in the Heyting algebra $\cal \mathfrak{H}$. A logician
would say that $\pi:\PL{S}\map\cal \mathfrak{H}$ is  an
\emph{$\mathfrak{H}$-valuation}, or \emph{$\mathfrak{H}$-model},
of the language $\PL{S}$.

Note that different systems, $S$, can have the same language. For
example, consider a point-particle moving in one dimension, with a
Hamiltonian $H=\frac{p^2} {2m}+V(x)$. Different potentials $V(x)$
correspond to different systems (in the sense in which we are
using the word `system'), but the physical quantities for these
systems---or, more precisely, the `names' of these quantities, for
example, `energy', `position', `momentum'---are the same for them
all. Consequently, the language $\PL{S}$ is independent of $V(x)$.
However, the \emph{representation} of, say, the proposition
``$H\varepsilon\De$'', with a specific subset of the state space
\emph{will} depend on the details of the Hamiltonian.

Clearly, a major consideration in using the language $\PL{S}$ is
choosing the Heyting algebra in which the representation takes
place. A fundamental result in topos theory is that the set of all
sub-objects of any object in a topos is  a Heyting algebra: these
are the Heyting algebras with which we will be concerned.

Of course, beyond  the language, ${\cal S}$, and its
representation $\pi$, lies the question of whether or not a
proposition is true. This requires the concept of a `state' which,
when specified, yields `truth values' for the primitive
propositions in $\PL{S}$. These are then extended recursively to
the rest of $\PL{S}$. In classical physics, the possible truth
values are just `true' or `false'. However,  the situation in
topos theory is more complex, and discussion is deferred to paper
II of the present series \cite{DI(2)}.

\paragraph{Introducing time dependence.} There is also
the question of `how things change in time'. In the form presented
above, the language $\PL{S}$ may seem geared towards a `canonical'
perspective in so far as the propositions concerned are,
presumably, to be asserted at a particular moment of time, and, as
such, deal with the values of physical quantities at that time. In
other words, the underlying spatio-temporal perspective seems
thoroughly `Newtonian'. This is partly true; but only partly,
since the phrase `physical quantity' can have meanings other than
the canonical one. For example, one could talk about the `time
average of momentum', and call that a physical quantity. In this
case, the propositions would be about \emph{histories} of the
system, not just `the way things are' at a particular moment in
time.

We will return to these extended versions  of the formalism in our
discussion of the higher-order language, $\L{S}$, in Section
\ref{SubSec:AdaptLanguage}. However, for the moment let us focus
on the canonical perspective, and the associated question of how
time dependence is to be incorporated. This can be addressed in
various ways.

One possibility is to attach a time label, $t$, to the physical
quantities, so that the primitive propositions  become of the form
``$A_t\,\varepsilon\,\De$''. In this case, the language itself
becomes time-dependent, so that we should write $\PL{S}_t$. One
might not like the idea of adding external labels in the language
and,  indeed, in our discussion of the higher-order language
$\L{S}$ we will strive to eliminate such things. However, in the
present case, in so far as $\De\subseteq\mathR$ is already an
`external' (to the language) entity, there seems no particular
objection to adding another one.

If we adopt this approach, the representation $\pi$ will map
``$A_t\,\varepsilon\,\De$'' to a time-dependent element,
$\pi(A_t\,\varepsilon\,\De)$,  of the Heyting algebra, $\cal
\mathfrak{H}$; one could say that this is a type of `Heisenberg
picture'. However, this suggests another option, which is to keep
the language time-independent, but allow the representation to be
time-dependent. In that case, $\pi_t(\Ain\De)$ will again be a
time-dependent member of $\mathfrak{H}$.

Another approach is to let the `truth object' in the theory be
time-dependent: this corresponds to a type of Schr\"odinger
picture. We will return to this subject in paper II\ where the
concept of a truth object is discussed in detail \cite{DI(2)}.

\subsubsection{The Representation of $\PL{S}$ in Classical Physics}
\label{SubSub:PhysRepPLS} Let us now look at the representation of
$\PL{S}$ that corresponds to classical physics. In this case, the
topos involved is just the category, $\Set$, of sets and functions
between sets.

We will denote by $\picl$ the representation of  $\PL{S}$ that
describes the classical, Hamiltonian mechanics of a system, $S$,
whose state-space is a symplectic (or Poisson) manifold $\S$. We
denote by $\breve{A}:\S\map\mathR$ the real-valued
function\footnote{In practice, $\breve{A}$ is required to be
measurable, or smooth, depending on the type of physical quantity
that $A$ is. However, for the most part, these details of
classical mechanics are not relevant to our discussions, and
usually we will not characterise $\breve{A}:\S\map\mathR$ beyond
just saying that it is a function/map from $\S$ to $\mathR$.} on
$\S$ that represents the physical quantity $A$.

Then the representation $\picl$ maps the primitive proposition
$\SAin\De $ to the subset of $\S$ given by
\begin{eqnarray}
\picl(\Ain\De)&:=&\{s\in\S\mid \breve{A}(s)\in\De\}
                                    \nonumber\\
                  &=& \breve{A}^{-1}(\De).\label{sigmaSAinDelta}
\end{eqnarray}
This representation can be extended to all the sentences in
$\PL{S}$ with the aid of \eqs{pi(a)}{pi(d)}. Note that, since
$\De$ is a Borel subset of $\mathR$, $\breve{A}^{-1}(\De)$ is a
Borel subset of the state-space $\S$. Hence, in this case,
 $\mathfrak{H}$ is equal to the Boolean algebra of all
Borel subsets of $\S$.

We note that, for all (Borel) subsets $\De_1,\De_2$ of $\mathR$ we
have
\begin{eqnarray}
   \breve{A}^{-1}(\De_1)\cap\breve{A}^{-1}(\De_2) &=& \breve{A}^{-1}(\De_1\cap\De_2)
                                                \label{A-1D1andD2}\\
   \breve{A}^{-1}(\De_1)\cup\breve{A}^{-1}(\De_2) &=& \breve{A}^{-1}(\De_1\cup\De_2)
                                                \label{A-1D1orD2}\\
                \neg\breve{A}^{-1}(\De_1)&=&\breve{A}^{-1}(\mathR\backslash\De_1)
                                                \label{notA-1D}
\end{eqnarray}
and hence all three conditions \eqs{AinD1andD2}{negAinD} that we
discussed earlier can be added consistently  to the language
$\PL{S}$ .

Consider now the assignment of  truth values to the propositions
in this theory. This involves the idea of a `state' which, in
classical physics, is simply an element $s$ of the state space
$\S$. Each state $s$ assigns to each primitive proposition
$\SAin\De$, a truth value, $\TVal{\Ain\De}{s}$,  which lies in the
set $\{{\rm false},{\rm true}\}$ (which we identify with
$\{0,1\}$) and is defined as\
\begin{equation}\label{Def:[AinD]Class}
        \TVal{\Ain\De}{s}:=
        \left\{\begin{array}{ll}
            1 & \mbox{\ if\ $\breve{A}(s)\in\De$;} \\
            0 & \mbox{\ otherwise}
         \end{array}
        \right.
\end{equation}
for all $s\in\S$.

\subsubsection{The Failure to Represent $\PL{S}$ in Standard Quantum Theory.}
The procedure above that works so easily for classical physics
fails completely if one tries to apply it  to standard quantum
theory.

In quantum physics, a physical quantity $A$ is represented by a
self-adjoint operator $\hat A$ on a Hilbert space $\Hi$, and the
proposition $\SAin\De$ is represented by the projection operator
$\hat E[A\in\De]$ which projects onto  the subset $\De$ of the
spectrum of $\hat A$; \ie\
\begin{equation}
        \pi(\Ain\De):=\hat E[A\in\De].
                        \label{Def:rhoQT}
\end{equation}

Of course, the set of all projection operators, $\PH$, in $\Hi$
has a `logic' of its own---the `quantum logic'\footnote{For an
excellent survey of quantum logic see \cite{DCG02}. This includes
a discussion of a first-order axiomatisation of quantum logic, and
with an associated sequent calculus. It is interesting to compare
our work with what the authors of this paper have done. We hope to
return to this at some time in the future.} of the Hilbert space
$\Hi$---but this is incompatible with the intuitionistic logic of
the language $\PL{S},$ and the representation \eq{Def:rhoQT}.

Indeed, since the `logic' $\PH$ is non-distributive, there will
exist non-commuting operators $\hat A,\hat B,\hat C$, and Borel
subsets  $\De_A,\De_B,\De_C$ of $\mathR$ such that\footnote{There
is a well-known example that uses three rays in $\mathR^2$, so
this phenomenon is not particularly exotic.}
\begin{eqnarray}
\hat E[A\in\De_A]\land\left(\hat E[B\in\De_B]
        \lor \hat E[C\in\De_C]\right)&\neq&\\
\left(\hat E[A\in\De_A]\land \hat E[B\in\De_B]\right)&\lor&
 \left(\hat E[A\in\De_A]\land\hat E[C\in\De_C]\right)
\end{eqnarray}
while, on the other hand, the logical bi-implication
\begin{equation}
        \alpha\land(\beta\lor\gamma)\Leftrightarrow
                (\alpha\land\beta)\lor(\alpha\land\gamma)
\end{equation}
can be deduced from the axioms of the language $\PL{S}$.

This failure of distributivity bars any na\"ive realist
interpretation of quantum logic. If an instrumentalist
interpretation is used instead, the spectral projectors $\hat
E[A\in\De]$ now represent propositions about what would happen
\emph{if} a measurement is made, not propositions about what is
`actually the case'.  And, of course, when a state is specified,
this does not yield actual truth values but only the Born-rule
probabilities of getting certain results.

\section{A Higher-Order, Typed  Language for Physics}
\label{Sec:TypedLanguage}
\subsection{The Basics of the Language $\L{S}$}
We want now  to consider the possibility of representing the
physical quantities of a system by arrows in a topos other than
$\Set$.

The physical meaning of such a quantity is not clear, \emph{a
priori}. Nor is it clear   \emph{what} it is that is being
represented in this way. However, what  \emph{is} clear is that in
such a situation it is no longer correct to work with a fixed
value-space $\mathR$. Rather, the target-object, $\R_S$, is
potentially topos-dependent, and therefore part of the
`representation'.

A powerful technique for allowing the quantity-value object to be
system-dependent is to add a symbol `$\R$' to the language.
Developing this line of thinking  suggests that `$\Si$', too,
should be added, as should a symbol `$A:\Si\map\R$', to be
construed as `what it is' that is represented by the arrow in a
topos. Similarly, there should be a symbol `$\O$', to act as the
linguistic precursor to the sub-object classifier in the topos; in
the topos $\Set$, this is just the set $\{0,1\}$.

The clean way of doing all this is to construct, what Bell
\cite{Bell88} calls, a `local language'. Our basic assumption is
that a unique local language, $\L{S}$, is associated with each
system $S$. Physical theories of $S$ then correspond to
representations of $\L{S}$ in appropriate topoi.

\paragraph{The symbols of $\L{S}$.} We first
consider  the minimal set of symbols needed to handle elementary
physics. For more sophisticated theories in physics, it will be
necessary to change, or enlarge, the set of `ground type' symbols.

The symbols for the local language, $\L{S}$, are defined
recursively as follows:
\begin{enumerate}
\item
\begin{enumerate}
  \item The basic \emph{type symbols} are $1,\O,\Si,\R$.
  The last two, $\Si$ and $\R$, are known as
  \emph{ground-type symbols}. They are the linguistic precursors
  of the state object, and quantity-value object, respectively.

If $T_1,T_2,\ldots,T_n$, $n\geq1$, are type symbols, then so
is\footnote{By definition, if $n=0$ then $T_1\times
T_2\times\cdots\times T_n:=1$.} $T_1\times T_2\times\cdots\times
T_n$.
        \item  If $T$ is a type symbol, then so is $PT$.
\end{enumerate}
\item
\begin{enumerate}
        \item   For each type symbol, $T$, there is associated a
        countable  set of \emph{variables of type $T$}.

        \item There is a special symbol $*$.
\end{enumerate}
\item
\begin{enumerate}
\item To each pair $(T_1,T_2)$ of type symbols there is associated
a set, $F_{\L{S}}(T_1,T_2)$, of \emph{function symbols}. Such a
symbol, $A$, is said to have \emph{signature} $T_1\map T_2$; this
is indicated  by writing $A:T_1\map T_2$.

\item Some of these sets of function symbols may be empty.
However, particular importance is attached to the set,
$F_{\L{S}}(\Si,\R)$, of   function symbols \mbox{$A:\Si\map\R$},
and we assume this set is non-empty.

\end{enumerate}
\end{enumerate}

The function symbols $A:\Si\map\R$ represent the `physical
quantities' of the system, and hence $F_{\L{S}}(\Si,\R)$ will
depend on the system. In fact, the only parts of the language that
are system-dependent are these function symbols.

For example, if $S_1$ is a point particle moving in one dimension,
the set of physical quantities could be chosen to be
$F_{\L{S_1}}(\Si,\R)=\{x,p,H\}$ which represent the position,
momentum, and energy of the system. On the other hand, if $S_2$ is
a particle moving in three dimensions, we could have
$F_{\L{S_2}}(\Si,{\cal R})=\{x,y,z,p_x,p_y,p_z,H\}$ to allow for
three-dimensional position and momentum. Or, we could decide to
add angular momentum too, to give the set $F_{\L{S_2}}(\Si,{\cal
R})=\{x,y,z,p_x,p_y,p_z,J_x,J_y,J_z,H\}$.

Note that, as with the propositional language $\PL{S}$, the fact
that a given system has a specific Hamiltonian\footnote{It must be
emphasised once more that the use of a local language is
\emph{not} restricted to standard, canonical systems in which the
concept of a `Hamiltonian' is meaningful. The scope of the
linguistic ideas is \emph{much} wider than that: the canonical
systems are only an example. Indeed, our long-term interest is in
the application of these ideas to quantum gravity, where the local
language is likely to be very different from that used here.
However, the basic  ideas are the same.}---expressed as a
particular function of position and momentum coordinates---is not
something that is to be coded into the language: instead, such
system dependence arises in the choice of \emph{representation} of
the language. This means that many different systems can have the
same local language.

Finally, it should be emphasised that this list of symbols is
minimal and one may want to add more. One obvious, general,
example is a type symbol $\mathN$, to be interpreted as the
linguistic analogue of the natural numbers. The language could
then be augmented with the axioms of Peano arithmetic.

\paragraph{The terms of ${\cal L}$.}
The next step is to enumerate the `terms' in the language,
together with their associated types \cite{Bell88,LamScott86}:
\begin{enumerate}
\item
\begin{enumerate}
\item For each type symbol $T$, the variables of type $T$ are
terms of type $T$.

\item The symbol $*$ is a term of type $1$.

\item A term of type $\O$ is called a \emph{formula}; a formula
with no free variables is called a \emph{sentence}.
\end{enumerate}

\item If $A$ is function symbol with signature $T_1\map T_2$, and
$t$ is a term of type $T_1$, then $A(t)$ is  term of type $T_2$.

In particular, if $A:\Si\map\R$ is a physical quantity, and $t$ is
a term of type $\Si$, then $A(t)$ is a term of type $\R$.

\item
\begin{enumerate}
        \item If $t_1,t_2,\ldots,t_n$ are terms of type
        $T_1,T_2,\ldots,T_n$, then
        $\langle t_1,t_2,\ldots,t_n\rangle$ is
        a term of type $T_1\times T_2\times\cdots\times T_n$.

       \item If $t$ is a term of type
       $T_1\times T_2\times\cdots\times T_n$,
       and if $1\leq i\leq n$, then $(t)_i$ is a term of type $T_i$.
\end{enumerate}

\item
\begin{enumerate}
        \item If $\omega$ is a term of type $\O$, and $\va{x}$ is a
        variable of type $T$, then $\{\va{x}\mid \omega\}$ is a term of
        type $PT$.

        \item If $t_1,t_2$ are terms of the same type, then $t_1=t_2$
        is a term of type $\O$.

        \item If $t_1,t_2$ are terms of type $T,PT$ respectively,
        then $t_1\in t_2$ is a term of type $\O$.
\end{enumerate}
\end{enumerate}

Note that the logical operations are not included in the set of
symbols. Instead,  they can all be defined using what is already
given. For example, (i) $true:= (*=*)$; and (ii) if $\alpha$ and
$\beta$ are terms of type $\O$, then\footnote{The parentheses
$(\;)$ are not  symbols in the language, they are just a way of
grouping letters and sentences.}
$\alpha\land\beta:=\big(\la\alpha, \beta\ra=\la{\rm true}, {\rm
true}\ra\big).$   Thus, in terms of the original set of symbols,
we have
\begin{equation}
  \alpha\land\beta:=\big(\langle\alpha,
   \beta\rangle=\langle*=*,*=*\rangle\big)
\end{equation}
and so on.

\paragraph{Terms of particular interest to us.}
Let  $A$ be a physical quantity in the set $\F{S}$, and therefore
a function symbol of signature $\Si\map\R$. In addition, let
$\va\De$ be a variable (and therefore a term) of type $P\R$; and
let $\va{s}$ be a variable (and therefore a term) of type $\Si$.
Then some terms of particular interest to us are  the following:
\begin{enumerate}
\item $A(\va{s})$ is a term of type $\R$ with a free variable,
$\va{s}$, of type $\Si$.

\item `$A(\va{s})\in\va\De$' is a term of type $\O$ with
free variables (i) $\va{s}$ of type $\Si$; and (ii) $\va{\De}$ of
type $P\R$.

\item $\{\va{s}\mid A(\va{s})\in\va\De\}$ is a term of
type $P\Si$ with a free variable $\va{\De}$ of type $P\R$.
\end{enumerate}
As we shall  see, $\{\va{s}\mid A(\va{s})\in\va\De\}$ and
`$A(\va{s})\in\va\De$' are (closely related) analogues of the
primitive propositions $\SAin\De$ in the propositional language
$\PL{S}$. However, there is a crucial difference. In  $\PL{S}$,
the `$\De$' in $\SAin\De$ is a specific subset of the external (to
the language) real line $\mathR$. On the other hand, in the local
language $\L{S}$, the `$\va\De$' in `$A(\va{s})\in\va\De$' is an
\emph{internal} variable within the language.

\paragraph{Adding axioms to the language.}
To make the language $\L{S}$ into  a deductive system we need to
add a set of appropriate axioms and rules of inference. The former
are expressed using \emph{sequents}: defined as expressions of the
form $\Ga:\alpha$ where $\alpha$ is a formula (a term of type
$\O$) and $\Ga$ is a set of such formula. The intention is that
`$\Ga:\alpha$' is to be read intuitively as ``the collection of
formula in $\Ga$ `imply' $\alpha$''. If $\Ga$ is empty we just
write $:\alpha$.

The basic axioms include things like `$\alpha:\alpha$'
(tautology), and  `$: \va{t}\in\{\va{t}\mid\alpha\}
\Leftrightarrow \alpha$' (comprehension) where $\va{t}$ is a
variable of type $T$. These axioms\footnote{The complete set is
\cite{Bell88}:
\begin{eqnarray*}
\mbox{Tautology:}       &&\alpha=\alpha\\[2pt]
\mbox{Unity}:           &&\va{x}_1=* \mbox{\ where $\va{x}_1$
                is a variable of type $1$.}       \\[2pt]
\mbox{Equality:}         && x=y,\alpha(\va{z}/x):
             \alpha(\va{z}/y). \mbox{ Here, $\alpha(\va{z}/x)$
             is the term $\alpha$ with $\va{z}$ replaced by the}\\
             &&\mbox{term $x$ for each free occurrence of the variable
         $\va{z}$. The terms $x$ and $y$ must}       \\
        && \mbox{be of the same type as $\va{z}$}.   \\[2pt]
\mbox{Products:}&& :(\la x_1,\ldots,x_n\ra)_i=x_i\\
                &&:x=\la(x)_1,\ldots,(x)_n\ra           \\[2pt]
\mbox{Comprehension:}&&:\va{t}\in\{\va{t}\mid\alpha\}
\Leftrightarrow \alpha
\end{eqnarray*}
} and the rules of inference (sophisticated analogues of
\emph{modus ponens})  give rise to a deductive system using
intuitionistic logic. For the details see
\cite{Bell88,LamScott86}.

However, for applications in physics we could add extra axioms (in
the form of sequents). For example, perhaps the quantity-value
object  should always be an abelian-group object\footnote{One
could go even further and add the axioms for real numbers. In this
case, in a representation of the language in a topos $\tau$, the
symbol $\R$ is mapped to the real-number object in the topos (if
there is one). However, the example of quantum theory suggests
that this is inappropriate \cite{DI(3)}.}? This can be coded into
the language by adding the axioms for an abelian group structure
for $\R$. This involves the following steps:
\begin{enumerate}
\item Add the following symbols:
\begin{enumerate}

\item A `unit' function symbol $0:1\map\R$; this will be the
linguistic analogue of the unit element in an abelian group.

\item An `addition' function symbol $+:\R\times\R\map\R$.

\item An `inverse' function symbol $-:\R\map\R$
\end{enumerate}

\item Then add axioms like
`$:\forall\va{r}\big(+\langle\va{r}, 0(*)\rangle= \va{r}\big)$'
where $\va{r}$ is a variable of type $\R$, and so on.
\end{enumerate}

For another example, consider a point particle moving in three
dimensions, with the function symbols
$F_{\L{S}}(\Si,{\R})=\{x,y,z,p_x,p_y,p_z,J_x,J_y,J_z,H\}$. As
$\L{S}$ stands,  there is no way to specify, for example, that
`$J_x=yp_z-zp_y$'. Such relations can only be implemented in a
\emph{representation} of the language. However, if
 this relation is felt to be `universal' (\ie\ it holds in all
physically-relevant representations) then it could be added to the
language with the use of extra axioms.

One of the delicate decisions that has to be made about $\L{S}$ is
what extra axioms  to add to the base language. Too few, and the
language lacks content; too many, and representations of potential
physical significance are excluded. This is one of the places in
the formalism where a degree of physical insight is necessary!

\subsection{Representing $\L{S}$ in a Topos}
The construction of a theory of the system $S$  involves choosing
a representation\footnote{The word `interpretation' is often used
in the mathematical literature, but we want to reserve that for
use in discussions of interpretations of quantum theory, and the
like.}/model, $\phi$, of the language $\L{S}$ in a
topos\footnote{A more comprehensive notation is $\tau_\phi(S)$,
which draws attention to the system $S$ under discussion;
similarly, the state object could be written as $\Sigma_{\phi,S}$,
and so on. This extended notation is used in paper IV where we are
concerned with the relations between \emph{different} systems, and
then it is essential to indicate which system is meant. However,
in the present paper, only one system at a time is being
considered, and so the truncated notation is fine.} $\tau_\phi$.
The choice of both topos and representation  depend on the
theory-type being used.

For example, consider a  system, $S$, that can be treated using
both classical physics and quantum physics, such as   a point
particle moving in three dimensions. Then, for the application of
the theory-type `classical physics', in a representation denoted
$\s$, the topos $\tau_\s$ is $\Set$, and $\Si$ is represented by
the symplectic manifold $\Si_\s:= T^*\mathR^3$.

On the other hand, for  the application of the theory-type
`quantum physics', $\tau_\phi$ is the topos, $\SetH{}$, of
presheaves over the category\footnote{We recall that the objects
in $\V{}$ are the unital, commutative von Neumann subalgebras of
the algebra, $\BH$, of all bounded operators on $\Hi$.} $\V{}$,
where ${\cal H}\simeq L^2(\mathR^3,d^3x)$ is the Hilbert space of
the system $S$. In this case, $\Si$ is represented by
$\Si_\phi:=\Sig$, where $\Sig$ is the spectral presheaf; this
representation is discussed at length in papers II and III
\cite{DI(2),DI(3)}. For both theory types, the \emph{details} of,
for example, the Hamiltonian, are coded in the representation.

We now list the $\tau_\phi$-representation of the  most
significant symbols and terms in our language, $\L{S}$ (we have
only picked out the parts that are immediately relevant to our
programme: for full details see \cite{Bell88, LamScott86}).
\begin{enumerate}

\item \begin{enumerate} \item The ground type symbols $\Si$ and
$\cal R$ are represented by objects $\Si_\phi$ and ${\cal R}_\phi$
in $\tau_\phi$. These are identified physically as the state
object, and quantity-value object, respectively.

   \item The symbol $\O$, is represented by
   $\O_\phi:=\O_{\tau_\phi}$, the sub-object classifier of
the topos $\tau_\phi$.

   \item The  symbol $1$, is represented by
$1_\phi:={1}_{\tau_\phi}$,
   the terminal object in $\tau_\phi$.
\end{enumerate}

\item For each type symbol $PT$, we have $(PT)_\phi:=PT_\phi$, the
power object of the object $T_\phi$ in $\tau_\phi$.

In particular, $(P\Si)_\phi=P\Si_\phi$ and $(P{\R})_\phi
=P\R_\phi$.

\item Each function symbol $A:\Si\map\R$ in $\F{S}$ (\ie\ each
physical quantity) is represented by an arrow
$A_\phi:\Si_\phi\map{\R}_\phi$ in $\tau_\phi$.

We will generally require the representation to be
\emph{faithful}: \ie\ the map $A\mapsto A_\phi$\ is one-to-one.

\item  A term of type $\O$ of the form
\q{$A(\va{s})\in\va\De$} (which has free variables
$\va{s},\va{\De}$ of type $\Si$ and $P\R$ respectively) is
represented by an arrow $\Val{A(\va{s})\in\va\De}_\phi
:\Si_\phi\times P{\R}_\phi\map \O_{\tau_\phi}$. In detail, this
arrow is
\begin{equation}
\Val{A(\va{s})\in\va\De}_\phi=e_{\R_\phi}\circ
        \la\Val{A(\va{s})}_\phi,\Val{\va\De}_\phi\ra
\end{equation}
where $e_{\R_\phi}:\R_\phi\times P\R_\phi\map\O_{\tau_\phi}$ is
the usual evaluation map;
$\Val{A(\va{s})}_\phi:\Si_\phi\map\R_\phi$ is the arrow $A_\phi$;
and $\Val{\va{\De}}_\phi:P\R_\phi\map P\R_\phi$ is the identity.

Thus $\Val{A(\va{s})\in\va\De}_\phi$ is the chain of arrows:
\begin{equation}
\Si_\phi\times P\R_\phi\mapright{A_\phi\times\id}
        \R_\phi\times P\R_\phi\mapright{e_{\R_\phi}}\O_{\tau_\phi}.
                                        \label{A(s)intildeDeChain}
\end{equation}
We see that the analogue of the `$\De$' used in the
$\PL{S}$-propositions $\SAin\De$ is played by sub-objects of
$\R_\phi$ (\ie\ global elements of $P\R_\phi$) in the domain of
the arrow in \eq{A(s)intildeDeChain}. These objects are, of
 course, representation-dependent (\ie\ they depend on $\phi$).

\item A term of type $P\Si$ of the form $\{\va{s}\mid
A(\va{s})\in\va\De\}$ (which has a free variable $\va\De$ of type
$P\R$) is represented by an arrow $\Val{\{\va{s}\mid
A(\va{s})\in\va\De\}}_\phi : P\R_\phi\map P\Si_\phi$. This arrow
is the power transpose\footnote{One of the basic properties of a
topos is that there is a one-to-one correspondence between arrows
$f:A\times B\map\O$ and arrows $\name{f}:A\map PB:=\O^B$. In
general, $\name{f}$ is called the \emph{power transpose} of $f$.
If $A\simeq 1$ then  $\name{f}$ is known as the \emph{name} of the
arrow $f:B\map\O$. See \eq{Def:exp} in the Appendix.} of
 $\Val{A(\va{s})\in\va\De}_\phi$:
\begin{equation}
\Val{\{\va{s}\mid A(\va{s})\in\va\De\}}_\phi =
\name{\Val{A(\va{s})\in\va\De}_\phi}\label{[]=nametildeDe}
\end{equation}

\item A term, $\omega$, of type $\O$ with no free variables is
represented by a global element $\Val{\omega}_\phi:1_{\tau_\phi}
\map \O_{\tau_\phi}$. These will typically act as `truth values'
for propositions about the system.

\item Any axioms that have been added to the language are required
to be represented by the arrow
$true:1_{\tau_\phi}\map\O_{\tau_\phi}$.
\end{enumerate}

\paragraph{The local set theory of a topos.}
We should emphasise that the decision to focus on the particular
type of language that we have, is not an arbitrary one. Indeed,
there is a deep connection between such languages and topos
theory.

In this context, we first note that to any local language, ${\cal
L}$, there is associated a `local set theory'. This involves
defining an `${\cal L}$-set' to be a term $X$ of power type (so
that expressions of the form $x\in X$ are meaningful) and with no
free variables. Analogues of all the usual set operations can be
defined on $\cal L$-sets. For example, if $X,Y$ are $\cal L$-sets
of type $PT$, one can define $X\cap Y:=\{\va{x}\mid \va{x}\in
X\land \va{x}\in Y\}$ where $\va{x}$ is a variable of type $T$.

Furthermore, each local set theory, ${\cal L}$, gives rise to an
associated topos, ${\cal C}({\cal L})$, whose objects are
equivalence classes of ${\cal L}$-sets, where $X\equiv Y$ is
defined to mean that the equation $X=Y$ (\ie\ a term of type
$\Omega$ with no free variables) can be proved using the sequent
calculus of the language with its axioms. From this perspective, a
representation of $\L{S}$ in a topos $\tau$ is equivalent to a
\emph{functor} from the topos ${\cal C}(\L{S})$ to $\tau$.

Conversely, for each topos $\tau$ there is a local language,
${\cal L}(\tau)$, whose ground-type  symbols are the objects of
$\tau$, and whose function symbols are the arrows in $\tau$. It
then follows that a representation of a local language, $\cal L$,
in $\tau$ is equivalent to a `translation' of ${\cal L}$ in ${\cal
L}(\tau)$.

Thus, a rather elegant way of summarising what is involved in
constructing a theory of physics is that we are \emph{translating}
the language, $\L{S}$, of the system in another local language,
$\L{\tau}$. As we will see in paper IV, the idea of translating
one local language into another plays a central role in the
discussion of composite systems and sub-systems \cite{DI(4)}.

\subsection{Classical Physics in the Local Language $\L{S}$}
The quantum theory representation of $\L{S}$ is studied in papers
II and III \cite{DI(2),DI(3)} of the present series. Here we will
look at the concrete form of the expressions in the previous
Section for the example of classical physics. In this case, for
all systems $S$, and all classical representations, $\s$, the
topos $\tau_\s$ is $\Set$. This representation of  $\L{S}$ has the
following ingredients:
\begin{enumerate}
\item
\begin{enumerate}
        \item The ground-type symbol $\Si$ is represented by
        a symplectic manifold, $\Si_\s$, that is the
        state-space for the system $S$.

        \item The ground-type symbol $\R$ is represented by the
        real line, \ie\ $\R_\s:=\mathR$.

        \item The type symbol $P\Si$ is represented by the set,
        $P\Si_\s$, of all subsets of the state space $\Si_\s$.

        The type symbol $P\R$ is represented by the set, $P\mathR$,
        of all subsets of $\mathR$.
\end{enumerate}

\item
\begin{enumerate}
   \item The  type symbol $\O$, is represented by
   $\Omega_\Set:=\{0,1\}$:  the sub-object classifier in $\Set$.

   \item The type symbol $1$, is represented by the
   singleton set, \ie\ $1_\Set=\{*\}$: the terminal object in $\Set$.
\end{enumerate}

\item Each function symbol $A:\Si\map\cal R$, and hence each
physical quantity, is represented by a real-valued function,
$A_\s:\Si_\s\map\mathR$, on the state space $\Si_\s$.

\item  The term \q{$A(\va{s})\in\va\De$} of type $\O$ (where
$\va{s}$ and $\va\De$ are free variables of type $\Si$ and $P\R$
respectively) is represented by the function
$\Val{A(\va{s})\in\va\De}_\s:\Si_\s\times P\mathR \map\{0,1\}$
that is defined by (c.f.\ \eq{A(s)intildeDeChain})
\begin{equation}
 \Val{A(\va{s})\in\va\De}_\s(s,\De)=
        \left\{\begin{array}{ll}
            1 & \mbox{\ if\ $A_\s(s)\in \De$;} \\
            0 & \mbox{\ otherwise.}
         \end{array}
        \right. \label{A(s)intildeDeChainCL}
\end{equation}
for all $(s,\De)\in\Si_\s\times P\mathR$.

\item The term  $\{\va{s}\mid A(\va{s}) \in\va{\De}\}$ of
type $P\Si$ (where $\va{\De}$ is a free variable of type $P\cal
R$) is represented by the function $\Val{\{\va{s}\mid A(\va{s})
\in\va{\De}\}}_\s: P\mathR \map P\Si_\s$ that is defined by
\begin{eqnarray}        \label{Def:sigmaS(A(s)inDelta)=}
        \Val{\{\va{s}\mid A(\va{s})
\in\va{\De}\}}_\s(\De) &:=&
        \{s\in\Si_\phi\mid A_\s(s)\in\De\}\nonumber\\
        &=& A_\s^{-1}(\De)
\end{eqnarray}
for all $\De\in P\mathR$.
\end{enumerate}

\subsection{Adapting the Language $\L{S}$ to Other Types of Physical
System}\label{SubSec:AdaptLanguage} Our central contention in this
series of papers is that (i) each physical system, $S$,  can be
equipped with a local language, $\L{S}$; and (ii) constructing an
explicit theory of $S$ in a particular theory-type is equivalent
to finding a representation of $\L{S}$ in a topos which may well
be other than the topos of sets.

There are many situations in which  the language is independent of
the theory-type, and then, for a given system $S$, the different
topos representations of $\L{S}$, correspond to the application of
the different theory-types to the same system $S$. We gave an
example earlier of a point particle moving in three dimensions:
the classical physics representation is in the topos $\Set$; and,
as shown in papers II\ and III, the quantum theory representation
is in the presheaf topos $\Set^{{\cal V}(L^2(\mathR^3,\,d^3x))}$ .

However, there are other situations where the relationship between
the language and its representations is more complicated than
this. In particular, there is the critical question about what
features of the theory should go into the language, and what into
the representation. Adding new features would begin by adding to,
or changing, the set  of ground-type symbols which generally
represent the entities that are going to be of generic interest
(such as a state object or quantity-value object). In doing this,
extra axioms may also be introduced to encode the properties that
the new objects are expected to possess in all the representations
that are of physical interest.

For example, suppose we want to use our formalism to discuss
space-time physics: where does the information about the
space-time go? If the subject is classical  field theory in a
curved  space-time, then the topos $\tau$ is  $\Set$, and the
space-time manifold is  part of the \emph{background} structure.
This makes it natural to have the manifold assumed in the
representation; \ie\ the information about the space-time is in
the representation.

However, alternatively one can add  a new ground type symbol,
`$M$', to the language, to serve as the linguistic progenitor of
`space-time'; thus $M$ would have  the same theoretical status as
the symbols $\Si$ and $\R$. A function symbol $\psi:M\map\R$ is
then  the progenitor of a physical field. In a representation
$\phi$, the object $M_\phi$ plays the role of `space-time' in the
topos $\tau_\phi$, and $\psi_\phi:M_\phi\map\R_\phi$ is the
representation of a field in this theory.

Of course, the language $\L{S}$  says nothing about what sort of
entity $M_\phi$  is, except in so far as such information is
encoded in extra axioms. For example, if the subject is classical
field theory, then $\tau_\phi=\Set$, and $M_\phi$ would be a
standard differentiable manifold. On the other hand, if the topos
$\tau_\phi$ admits `infinitesimals', then $M_\phi$ could be a
manifold according to the language of synthetic differential
geometry \cite{Kock81}.

\emph{A fortiori}, the same type of argument applies to the status
of `time' in a canonical theory. In particular, it is possible to
add a ground type symbol, $\typeTime$, so that, in any
representation, $\phi$, the object $\typeTime_\phi$ in the topos
$\tau_\phi$ is the analogue of the `time-line' for that theory.
For standard physics in $\Set$ we have $\typeTime_\phi=\mathR$,
but the form of $\typeTime_\phi$ in a more general topos,
$\tau_\phi$, would be a rich subject for speculation.

The addition of  a  `time-type' symbol, $\typeTime$, to the
language $\L{S}$ is a prime example of a situation where  one
might want to add extra axioms. These could involve ordering
properties, or algebraic  properties like those of  an abelian
group, and so on. These properties would  be realised in any
representation as the corresponding type of object in the topos
$\tau_\phi$. Thus abelian group axioms mean that $\typeTime_\phi$
is an abelian-group object in $\tau_\phi$; total-ordering axioms
for the time-type $\typeTime$ mean that $\typeTime_\phi$ is a
totally-ordered object in $\tau_\phi$, and so on.

As a rather interesting extension of this idea, one could have a
space-time ground type symbol $M$, but then add the axioms for a
partial ordering. In that case, $M_\phi$ would be a poset-object
in $\tau_\phi$, which could be interpreted physically as the
$\tau_\phi$-analogue of a causal set \cite{Dow05}.

Yet another possibility is to develop a language for history
theories, and use it study the topos version of the
consistent-histories approach to quantum theory.

We will return to some of these ideas in future publications.

\section{Conclusion}\label{Sec:Concl}
In this paper, the first in a series, we have introduced the idea
that a formal language can be attached to each physical system,
and that constructing a theory of that system is equivalent to
finding a representation of this  language in an appropriate
topos. The long-term goal of this research programme is to provide
a novel framework for constructing theories of physics in general;
in particular, to construct theories that go `beyond' standard
quantum theory, and especially in the direction of quantum
cosmology. In doing so, we have constructed a formalism that is
not tied to the familiar use of Hilbert spaces, or formal path
integrals, and which, therefore, need not assume \emph{a priori}
the use of continuum quantities in physics.

We have introduced two different types of language that can apply
to a given system $S$. The first is the propositional language,
$\PL{S}$, that deals only with propositions of the form
$\SAin\De$. The intention is represent these propositions in a
Heyting algebra of sub-objects of some object in a topos that is
identified as the analogue of a `state space'. The simplest
example is classical physics, where propositions are represented
by  the Boolean algebra of (Borel) subsets of the classical state
space. The example of quantum theory is considerably more
interesting and is discussed in detail in paper II \cite{DI(2)}.

The second type of language that we discussed is considerably more
powerful. This is the `local' language $\L{S}$ which includes
symbols for the state object and quantity-value object (and/or
whatever theoretical entities are felt to be of
representation-independent importance), as well as symbols for the
physical quantities in the system. The key idea is that
constructing a theory of $S$ is equivalent is to finding a
representation of this entire language (not just the propositional
part) in a topos. As with $\PL{S}$, the language $\L{S}$ forms a
deductive system that is based on intuitionistic logic: something
that is naturally adapted to finding a representation in a topos.

Any  theory of this type is necessarily `neo-realist' in the sense
that physical quantities are represented by arrows
$A_\phi:\Si_\phi\map\R_\phi$; and propositions are represented by
sub-objects of $\Si_\phi$, the set  of which is a Heyting algebra.
In this sense, these topos-based theories all `look' like
classical physics, except of course that, generally speaking, the
topos concerned is not $\Set$.

\bigskip
\noindent {\bf Acknowledgements} This research was supported by
grant RFP1-06-04 from The Foundational Questions Institute
(fqxi.org).  AD gratefully acknowledges financial support from the
DAAD.

This work is also supported in part by the EC Marie Curie Research
and Training Network ``ENRAGE'' (European Network on Random
Geometry) MRTN-CT-2004-005616.

We thank Jeremy Butterfield for a careful reading of the final
draft of this paper.

\appendix
\section{A Brief Account of the Relevant Parts of Topos Theory}
\label{Sec:mathprel}

\subsection{Presheaves on a Poset}
\label{SubSec:presheaves-poset} Topos theory is a remarkably rich
branch of mathematics which can be approached from a variety of
different viewpoints. The basic area of mathematics  is category
theory; where, we recall, a category consists of a collection of
\emph{objects} and a collection of \emph{morphisms} (or
\emph{arrows}).

In the special case of the category of sets, the objects are sets,
and a morphism is a function between a pair of sets. In general,
each morphism $f$ in a category is associated with a pair of
objects, known as its `domain' and  `codomain', and is written as
$f:B\map A$ where $B$ and $A$ are the domain and codomain
respectively. Note that this arrow notation is used even if $f$ is
not a function in the normal set-theoretic sense. A key ingredient
in the definition of a category is that if $f:B\map A$ and
$g:C\map B$ (\ie\ the codomain of $g$ is equal to the domain of
$f$) then $f$ and $g$ can be `composed' to give an arrow $f\circ
g:C\map A$; in the case of the category of sets, this is just the
usual composition of functions.

A simple  example of a category is given by  any partially-ordered
set (`poset') $\cal C$: (i) the objects are defined to be the
elements of $\cal C$; and (ii) if $p,q\in\cal C$, a morphism from
$p$ to $q$ is defined to exist if, and only if, $p\preceq q$ in
the poset structure.  Thus, in a poset regarded as a category,
there is at most one morphism between any pair of objects
$p,q\in\cal C$; if it exists, we shall write this morphism as
$i_{pq}:p\map q$. This example is important for us in form of the
`category of contexts', $\V{}$, in quantum theory (see papers
II-IV). The objects in $\V{}$ are the commutative, unital von
Neumann subalgebras of the algebra, $\BH$, of all bounded
operators on the Hilbert space $\Hi$. (Unital means that all these
algebras contain the identity operator $\hat 1\in\BH$.)

\paragraph{The definition of a topos.}From our perspective,
the most relevant feature of a topos, $\tau$,  is that it is a
category which behaves in many ways like the category of sets
\cite{Gol84,MM92}. Most of the precise details are not necessary
for the present series of papers, but here we will list some of
the most important ones for our purposes:
\begin{enumerate}
\item There is a terminal object $1_\tau$ in $\tau$; this means
that given any object $A$ in the topos, there is a unique arrow
$A\map 1_\tau$.

For any object $A$ in the topos, an arrow $1_\tau\map A$ is called
a \emph{global element}\footnote{In the category of sets, $\Set$,
the terminal object $1_\Set$ is a singleton set $\{*\}$. It
follows that the elements of $\Ga A$\ are in one-to-one
correspondence with the elements of $A$. } of $A$. The set of all
global elements of $A$ is denoted $\Ga A$.

Given $A,B\in\Ob{\tau}$, there is a product $A\times B$ in $\tau$.
In fact, a topos always has \emph{pull-backs}, and the product is
just a special case of this.\footnote{The conditions in 1.\ above
are equivalent to saying that $\tau$ is \emph{finitely complete}.}

\item There is an initial object $0_\tau$ in $\tau$. This means
that given any object $A$ in the topos, there is a unique arrow
$0_\tau\map A$.

Given $A,B\in\Ob{\tau}$, there is a co-product $A\sqcup B$ in
$\tau$.  In fact, a topos always has \emph{push-outs}, and the
co-product is just a special case of this.\footnote{The conditions
in 2.\ above are equivalent to saying that $\tau$ is
\emph{finitely co-complete}.}

\item There is \emph{exponentiation}: \ie\ given objects $A,B$ in
$\tau$ we can form the object $A^B$, which is the topos analogue
of the set of functions from $B$ to $A$ in set theory. The
definitive property of exponentiation is that, given any object
$C$, there is an isomorphism
\begin{equation}
\Hom{\tau}{C}{A^B}\simeq \Hom{\tau}{C\times B}{A}\label{Def:exp}
\end{equation}
that is natural in $A$ and $C$.

\item There is a sub-object classifier $\O_\tau$.
\end{enumerate}

The last item is of particular importance to us as it is the
source of the Heyting algebras that we use so much. To explain
what is meant, let us first consider the familiar topos, $\Set$,
of sets. There, the subsets $K\subseteq X$ of a set $X$ are in
one-to-one correspondence with functions $\cha{K}:X\map\{0,1\}$,
where $\cha{K}(x)=1$ if $x\in K$, and $\cha{K}(x)=0$ otherwise.
Thus the target space $\{0,1\}$ can be regarded as the simplest
`false-true' Boolean algebra, and the mathematical proposition
``$x\in K$'' is true if $\cha{K}(x)=1$, and false otherwise.

In the case of a topos, $\tau$, the sub-objects\footnote{An object
$K$ is a \emph{sub-object} of another object $X$ if there is a
monic arrow $K\hookrightarrow X$. In the topos $\Set$ of sets,
this is equivalent to saying that $K$ is a subset of $X$.}  $K$ of
an object $X$ in the topos are in one-to-one correspondence with
arrows $\cha{K}:X\map \O_\tau$, where the special object
$\O_\tau$---called the `sub-object classifier', or `object of
truth values'---plays an analogous role to that of $\{0,1\}$ in
the category of sets.

An important property for us is that, in any topos $\tau$, the
collection, $\Sub{A}$, of sub-objects of an object $A$ forms a
\emph{Heyting algebra}. The reader is referred to the standard
texts for proofs (for example, see \cite{Gol84}, p151).

\paragraph{The idea of a presheaf.} To illustrate the main
ideas, we will first give a few definitions from the theory of
presheaves on a partially ordered set (or `poset'); in the case of
quantum theory, this poset is the space of `contexts' in which
propositions are asserted. We shall then use these ideas to
motivate the definition of a presheaf on a general category.  Only
the briefest of treatments is given here, and the reader is
referred to the standard literature for more information
\cite{Gol84,MM92}.

A \emph{presheaf} (also known as a \emph{varying set\/}) $\ps{X}$
on a poset $\cal C$ is a function that assigns to each $p\in\cal
C$, a set $\ps{X}_p$; and to each pair $p\preceq q$ (\ie\
$i_{pq}:p\map q$), a map $\ps{X}_{qp}:\ps{X}_q\map \ps{X}_p$ such
that (i) $\ps{X}_{pp}:\ps{X}_p\map\ps{X}_p$ is the identity map
${\rm id}_{{\ps{X}_p}}$ on $\ps{X}_p$, and (ii) whenever $p\preceq
q\preceq r$, the composite map
$\ps{X}_r\stackrel{\ps{X}_{rq}}\longrightarrow
\ps{X}_q\stackrel{\ps{X}_{qp}}\longrightarrow \ps{X}_p$ is equal
to
 $\ps{X}_r\stackrel{\ps{X}_{rp}}\longrightarrow \ps{X}_p$, so that
\begin{equation}
        \ps{X}_{rp}= \ps{X}_{qp}\circ\ps{X}_{rq}. \label{Xrp=XqpXrq}
\end{equation}
The notation $\ps{X}_{qp}$ is shorthand for the more cumbersome
$\ps{X}(i_{pq})$; see below in the definition of a functor.

An \emph{arrow}, or \emph{natural transformation} $\eta:\ps{X}\map
\ps{Y}$ between two presheaves $\ps{X},\ps{Y}$ on $\cal C$ is a
family of maps $\eta_p:\ps{X}_p\map \ps{Y}_p$, $p\in\cal C$, that
satisfy the intertwining conditions
\begin{equation}
        \eta_p\circ\ps{X}_{qp}=\ps{Y}_{qp}\circ\eta_q
\end{equation}
whenever $p\preceq q$. This is equivalent to the commutative
diagram
\begin{equation}                                                                \label{Def:eta}
                \setsqparms[1`1`1`1;1000`700]
    \square[\ps{X}_q`\ps{X}_p`\ps{Y}_q`\ps{Y}_p;
    \ps{X}_{qp}`\eta_q`\eta_p`\ps{Y}_{qp}]
\end{equation}

A \emph{sub-object} of a presheaf $\ps{X}$ is a presheaf $\ps{K}$,
with an arrow  $i:\ps{K}\map \ps{X}$ such that (i)
$\ps{K}_p\subseteq \ps{X}_p$ for all $p\in\cal C$; and (ii) for
all $p\preceq q$, the map $K_{qp}:\ps{K}_q\map \ps{K}_p$ is the
restriction of $\ps{X}_{qp}:\ps{X}_q\map \ps{X}_p$ to the subset
$\ps{K}_q\subseteq\ps{X}_q$. This is shown in the commutative
diagram
\begin{equation}                                                                \label{cd}
                \setsqparms[1`1`1`1;1000`700]
    \square[\ps{K}_q`\ps{K}_p`\ps{X}_q`\ps{X}_p;
    \ps{K}_{qp}```\ps{X}_{qp}]
\end{equation}
where the vertical arrows are subset inclusions.

The collection of all presheaves on a poset $\cal C$ forms a
category, denoted $\SetC{\cal C}$.  The arrows/morphisms between
presheaves in this category are defined as the arrows above.

\subsection{Presheaves on a General Category}
\label{SubSec:presheaves-gen-cat} The ideas sketched above admit
an immediate generalization to the theory of presheaves on an
arbitrary `small' category $\cal C$ (the qualification `small'
means that the collection of objects is a genuine set, as is the
collection of all arrows/morphisms between any pair of objects).
To make the necessary definition we first need the idea of a
`functor':

\paragraph*{1. The idea of a functor:}
A central concept is that of a `functor' between a pair of
categories $\cal C$ and $\cal D$. Broadly speaking, this is an
arrow-preserving function from one category to the other. The
precise definition is as follows.

\begin{definition}
\begin{enumerate}

\item {A \emph{covariant functor} $\fu{F}$ from a category $\cal
C$ to a category $\cal D$ is a function that assigns
    \begin{enumerate}
        \item to each $\cal C$-object $A$, a $\cal D$-object
        $\fu{F}_A$;

        \item {to each $\cal C$-morphism $f:B\map A$, a
$\cal D$-morphism $\fu{F}(f):\fu{F}_B\map \fu{F}_A$ such that
$\fu{F}(\id_A)={\rm id}_{\fu{F}_A}$; and, if $g:C\map B$, and
$f:B\map A$ then
    \begin{equation}
        \fu{F}(f\circ g)=\fu{F}(f)\circ
                \fu{F}(g).     \label{Def:covfunct}
    \end{equation}
        }
    \end{enumerate}
    }

\item {A {\em contravariant functor\/} $\fu{X}$ from a category
$\cal C$ to a category $\cal D$ is a function that assigns
\begin{enumerate} \item to each $\cal C$-object $A$, a $\cal
D$-object $\fu{X}_A$;

    \item {to each $\cal C$-morphism $f:B\map A$, a $\cal
D$-morphism $\fu{X}(f):\fu{X}_A\map \fu{X}_B$ such that
$\fu{X}(\id_A)=\id_{\fu{X}_A}$; and, if $g:C\map B$, and $f:B\map
A$ then
    \begin{equation}
        \fu{X}(f\circ g)=\fu{X}(g)\circ\fu{X}(f).
                        \label{Def:confunct}
    \end{equation}
        }
    \end{enumerate}
    }
\end{enumerate}
\end{definition}

The connection with the idea of a presheaf on a poset is
straightforward. As mentioned above, a poset $\cal C$ can be
regarded as a category in its own right, and it is clear that a
presheaf on the poset $\cal C$ is the same thing as a
contravariant functor $\ps{X}$ from the category $\cal C$ to the
category $\Set$ of normal sets. Equivalently, it is a covariant
functor from the `opposite' category\footnote{The `opposite' of a
category $\cal C$ is a category, denoted ${\cal C}^\op$, whose
objects are the same as those of $\cal C$, and whose morphisms are
defined to be the opposite of those of $\cal C$; \ie\ a morphism
$f:A\map B$ in ${\cal C}^\op$ is said to exist if, and only if,
there is a morphism $f:B\map A$ in $\cal C$.} ${\cal C}^{\rm op}$
to $\Set$. Clearly,  \eq{Xrp=XqpXrq} corresponds to the
contravariant condition  \eq{Def:confunct}. Note that
mathematicians usually call the objects in $\cal C$ `stages of
truth', or just `stages'. For us they are `contexts'.

\paragraph*{2. Presheaves on an arbitrary category $\cal C$:}
These remarks motivate the definition of a presheaf on an
arbitrary small category $\cal C$: namely, a {\em presheaf\/} on
$\cal C$ is a covariant functor\footnote{Throughout this series of
papers, a presheaf is indicated by a letter that is underlined.}
$\ps{X}:{\cal C}^\op\map\Set$ from ${\cal C}^\op$ to the category
of sets. Equivalently, a presheaf is a contravariant functor from
$\cal C$ to the category of sets.

We want to make the collection of presheaves on $\cal C$ into a
category, and therefore we need to define what is meant by a
`morphism' between two presheaves $\ps{X}$ and $\ps{Y}$.  The
intuitive idea is that such a morphism from $\ps{X}$ to $\ps{Y}$
must give a `picture' of $\ps{X}$ within $\ps{Y}$. Formally, such
a morphism is defined to be a \emph{natural transformation}
$N:\ps{X}\map\ps{Y}$, by which is meant a family of maps (called
the \emph{components} of $N$) $N_A:\ps{X}_A\map\ps{Y}_A$,
$A\in\Ob{\cal C}$, such that if $f:B\map A$ is a morphism in $\cal
C$, then the composite map $\ps{X}_{A}
\stackrel{N_A}\longrightarrow\ps{Y}_A\stackrel{\ps{Y}(f)}
\longrightarrow\ps{Y}_B$ is equal to $\ps{X}_A
\stackrel{\ps{X}(f)}\longrightarrow\ps{X}_B\stackrel{N_B}
\longrightarrow \ps{Y}_A$. In other words, we have the commutative
diagram
\begin{equation}                                                                \label{cdNT}
                \setsqparms[1`1`1`1;1000`700]
    \square[\ps{X}_A`\ps{X}_B`\ps{Y}_A`\ps{Y}_B;
    \ps{X}(f)`N_A`N_B`\ps{Y}(f)]
\end{equation}
of which \eq{Def:eta} is clearly a special case. The category of
presheaves on $\cal C$ equipped with these morphisms is denoted
$\SetC{\cal C}$.

The idea of a sub-object generalizes in an obvious way. Thus we
say that $\ps{K}$ is a \emph{sub-object} of $\ps{X}$ if there is a
morphism in the category of presheaves (\ie\ a natural
transformation) $\iota:\ps{K}\map\ps{X}$ with the property that,
for each $A$, the component map $\iota_A:\ps{K}_A\map\ps{X}_A$ is
a subset embedding, \ie\ $\ps{K}_A\subseteq \ps{X}_A$. Thus, if
$f:B\map A$ is any morphism in $\cal C$, we get the analogue of
the commutative diagram \eq{cd}:
\begin{equation}                                                                \label{subobject}
                \setsqparms[1`1`1`1;1000`700]
    \square[\ps{K}_A`\ps{K}_B`\ps{X}_A`\ps{X}_B;
    \ps{K}(f)```\ps{X}(f)]
\end{equation}
where, once again, the vertical arrows are subset inclusions.

The category of presheaves on $\cal C$, $\Set^{{\cal C}^{\rm
op}}$, forms a topos. We do not need the full definition of a
topos; but we do need the idea, mentioned in Section
\ref{SubSec:presheaves-poset}, that a topos has a sub-object
classifier $\O$, to which we now turn.

\paragraph*{3. Sieves and the sub-object classifier $\Om$.}
Among the key concepts in presheaf theory is that of a `sieve',
which plays a central role in the construction of the sub-object
classifier in the topos of presheaves on a category $\cal C$.

A {\em sieve\/} on an object $A$ in $\cal C$ is defined to be a
collection $S$ of morphisms $f:B\map A$ in $\cal C$ with the
property that if $f:B\map A$ belongs to $S$, and if $g:C\map B$ is
any morphism with co-domain $B$, then $f\circ g:C\map A$ also
belongs to $S$. In the simple case where $\cal C$ is a poset, a
sieve on $p\in\cal C$ is any subset $S$ of $\cal C$ such that if
$r\in S$ then (i) $r\preceq p$, and (ii) $r'\in S$ for all
$r'\preceq r$; in other words, a sieve is nothing but a {\em
lower\/} set in the poset.

The presheaf $\Om:{\cal C}\map \Set$ is now defined as follows. If
$A$ is an object in $\cal C$, then $\Om_A$ is defined to be the
set of all sieves on $A$; and if $f:B\map A$, then
$\Om(f):\Om_A\map\Om_B$ is defined as
\begin{equation}
{\ps{\O}}(f)(S):= \{h:C\map B\mid f\circ h\in S\}
                                \label{Def:Om(f)}
\end{equation}
for all $S\in\Om_A$; the sieve $\Om(f)(S)$ is often written as
$f^*(S)$, and is known as the {\em pull-back\/} to $B$ of the
sieve $S$ on $A$ by the morphism $f:B\map A$.

It should be noted that if $S$ is a sieve on $A$, and if $f:B\map
A$ belongs to $S$, then from the defining property of a sieve we
have
\begin{equation}
        f^*(S):=\{h:C\map B\mid f\circ h\in S\}=
\{h:C\map B\}=:\ \downarrow\!\!B     \label{f*S}
\end{equation}
where $\downarrow\!\!B$ denotes the {\em principal\/} sieve on
$B$, defined to be the set of all morphisms in $\cal C$ whose
codomain is $B$. In words: the pull-back of any sieve on $A$ by a
morphism from $B$ to $A$ that belongs to the sieve, is the {\em
principal\/} sieve on $B$.

If $\cal C$ is a poset, the pull-back operation corresponds to a
family of maps $\Om_{qp}:\Om_q\map\Om_p$ (where $\Om_p$ denotes
the set of all sieves/lower sets on $p$ in the poset) defined by
$\Om_{qp}=\Om(i_{pq})$ if $i_{pq}:p\map q$ ({\em i.e.}, $p\preceq
q$). It is straightforward to check that if $S\in\Om_q$, then
\begin{equation}
\Om_{qp}(S):=\downarrow\!{p}\cap S \label{Def:Omqp}
\end{equation}
where $\downarrow\!{p}:=\{r\in{\cal C}\mid r\preceq p\}$.

A crucial property of sieves is that the set $\Om_A$ of sieves on
$A$ has the structure of a Heyting algebra. Specifically, $\Om_A$
is a Heyting algebra where the unit element $1_{\Om_A}$ in $\Om_A$
is the principal sieve $\downarrow\!\!A$, and the null element
$0_{\Om_A}$ is the empty sieve $\emptyset$. The partial ordering
in $\Om_A$ is defined by $S_1\preceq S_2$ if, and only if,
$S_1\subseteq S_2$; and the logical connectives are defined as:
\begin{eqnarray}
    && S_1\land S_2:=S_1\cap S_2    \label{Def:S1landS2}\\
    && S_1\lor S_2:=S_1\cup S_2     \label{Def:S1lorS2} \\
    &&S_1\Rightarrow S_2:=\{f:B\map A\mid
    \mbox{ $\forall$ $g:C\map B$ if $f\circ g\in S_1$ then
                $f\circ g\in S_2$}\}
\end{eqnarray}
As in any Heyting algebra, the negation of an element $S$ (called
the {\em pseudo-complement\/} of $S$) is defined as $\neg
S:=S\Rightarrow 0$; so that
\begin{equation}
    \neg S:=\{f:B\map A\mid \mbox{for all
$g:C\map B$, $f\circ g\not\in S$} \}.    \label{Def:negS}
\end{equation}

It can be shown that the presheaf $\Om$ is a sub-object classifier
for the topos $\SetC{\cal C}$. That is to say, sub-objects of any
object $\ps{X}$ in this topos ({\em i.e.}, any presheaf on $\cal
C$) are in one-to-one correspondence with morphisms
$\chi:\ps{X}\map {\ps{\O}}$. This works as follows. First, let
$\ps K$ be a sub-object of $\ps{X}$.  Then there is an associated
{\em characteristic\/} morphism
$\chi_{\ps{K}}:\ps{X}\map{\ps{\O}}$, whose `component'
$\chi_{\ps{K}A}:\ps{X}_A\map\Om_A$ at each stage/context $A$ in
$\cal C$ is defined as
\begin{equation}
    \chi_{\ps{K} A}(x):=\{f:B\map A\mid \ps{X}(f)(x)\in
                                \ps{K}_B\} \label{Def:chiKA}
\end{equation}
for all $x\in \ps{X}_A$. That the right hand side of
\eq{Def:chiKA} actually {\em is\/} a sieve on $A$ follows from the
defining properties of a sub-object.

Thus, in each `branch' of the category $\cal C$ going `down' from
the stage $A$, $\cha{\ps{K}}{}_A(x)$ picks out the first member
$B$ in that branch for which $\ps{X}(f)(x)$ lies in the subset
$\ps{K}_B$, and the commutative diagram \eq{subobject} then
guarantees that $\ps{X}(h\circ f)(x)$ will lie in $\ps{K}_C$ for
all $h:C\map B$.  Thus each stage  $A$ in $\cal C$ serves as a
possible context for an assignment to each $x\in \ps{X}_A$ of a
generalised truth value---a sieve belonging to the Heyting algebra
$\Om_A$.  This is the sense in which contextual, generalised truth
values arise naturally in a topos of presheaves.

There is a converse to \eq{Def:chiKA}: namely, each morphism
$\chi:\ps{X}\map{\ps{\O}}$ ({\em i.e.}, a natural transformation
between the presheaves $\ps{X}$ and ${\ps{\O}}$) defines a
sub-object $\ps{K}^\chi$ of $\ps{X}$ via
\begin{equation}
    \ps{K}^\chi_A:=\chi_A^{-1}\{1_{\Om_A}\}.
                            \label{Def:KchiA}
\end{equation}
at each stage $A$.

\paragraph*{4. Global elements of a presheaf:}
We recall that, in any topos, $\tau$, a \emph{terminal object} is
defined to be an object $1_\tau$ with the property that, for any
object $X$ in the category, there is a unique morphism $X\map
1_\tau$; it is easy to show that terminal objects are unique up to
isomorphism. A \emph{global element} of an object $X$ is then
defined to be any morphism $s:1_\tau\map X$. The motivation for
this nomenclature is that, in the case of the category of sets, a
terminal object is any singleton set $\{*\}$; and then it is true
that there is a one-to-one correspondence between the elements of
a set $X$ and functions from $\{*\}$ to $X$.

For the category of presheaves on $\cal C$, a terminal object
$\ps{1}:{\cal C}\map \Set$ can be defined by $\ps{1}_A:=\{*\}$ at
all stages $A$ in $\cal C$; if $f:B\map A$ is a morphism in $\cal
C$ then $\ps{1}(f):\{*\}\map\{*\}$ is defined to be the map
$*\mapsto *$. This is indeed a terminal object since, for any
presheaf $\ps{X}$, we can define a unique natural transformation
$N:\ps{X}\map\ps{1}$ whose components
$N_A:\ps{X}(A)\map\ps{1}_A=\{*\}$ are the constant maps $x\mapsto
*$ for all $x\in\ps{X}_A$.

A global element of a presheaf $\ps{X}$ is also called a
\emph{global section}. As a morphism $\ga:\ps{1}\map\ps{X}$ in the
topos $\SetC{\cal C}$, a global element corresponds to a choice of
an element $\ga_A\in\ps{X}_A$ for each stage  $A$ in $\cal C$,
such that, if $f:B\map A$, the `matching condition'
\begin{equation}
    \ps{X}(f)(\ga_A)=\ga_B \label{Def:global}
\end{equation}
is satisfied.

\end{document}